\documentclass[twocolumn]{autart}
\input epsf
\usepackage{cite,epsfig}
\usepackage{graphicx,amsmath,amssymb,color}
\usepackage{algpseudocode}

\usepackage{calc}
\usepackage[justification=centering]{caption}
\usepackage{wrapfig}
\usepackage{diagbox}
\usepackage{subcaption}
\usepackage{algorithm}
\usepackage{algpseudocode}
\newcommand{\zc}{}
\newcommand{\mz}{\color{blue}}

\graphicspath{ {./Images-PaillierSecondorderTACTemplate/} }
\newtheorem{theorem}{Theorem}
\newtheorem{definition}{Definition}
\newtheorem{lemma}{Lemma}

\newtheorem{remark}{Remark}

\newcommand{\eproof}{\hfill\rule{2mm}{2mm}}

\begin{document}

\begin{frontmatter}

\title{Secure and Privacy Preserving Consensus for Second-order Systems Based on Paillier Encryption
\thanksref{footnoteinfo}}

\thanks[footnoteinfo]{This paper was not presented at any IFAC
meeting. The work
was substantially supported by the Australian Research Council
under grant No. DP150103745.}

\author{Wentuo Fang}\ead{Wentuo.fang@uon.edu.au},  
\author{Mohsen Zamani}\ead{mohsen.zamani@newcastle.edu.au}, and
\author{Zhiyong Chen}\ead{zhiyong.chen@newcastle.edu.au}

\address{School of Electrical Engineering and Computing,
The University of Newcastle, Callaghan, NSW 2308, Australia\\
Tel: +61 2 4921 6352, Fax: +61 2 4921 6993}  


\begin{abstract}
This paper aims at
secure and privacy preserving consensus algorithms of networked systems. 
Due to the technical challenges behind decentralized design of such algorithms, the existing results are mainly restricted to a network of  systems with simplest first-order dynamics. 
Like many other control problems, breakthrough of the gap 
between first-order dynamics and higher-order ones demands for more advanced technical developments. 
In this paper, we explore a Paillier encryption based average consensus algorithm for a network of  systems with second-order dynamics, with randomness added to network weights. 
The conditions for privacy preserving, especially depending on consensus rate, are thoroughly studied with theoretical analysis and numerical verification. 
\end{abstract}

\end{frontmatter}

\section{Introduction}
	
{\zc  In recent years research on network security has attracted great attention and significant progresses have been made in this field. For example, cyber security of power systems, especially on smart grid, is one of the most popular topics; see, e.g.,  \cite{wang2013cyber,yan2012survey,liu2012cyber}. Network security has also be deeply investigated in many other applications including  but not limited to cyber security management of 
industrial systems \cite{knowles2015survey}, secure control for resource-limited adversaries \cite{teixeira2015secure}, relay attacks \cite{mo2009secure}  and  cloud computing \cite{krutz2010cloud}. 
There are various techniques for keeping privacy and security of network systems, among which encryption is an effective but challenging method. A typical encryption technique involving trusted third-party can be found in many references such as \cite{lopez2012fly,brenner2011secret,ren2012security,kerschbaum2012outsourced,shoukry2016privacy}. In particular, there has been a great impetus within systems and control community to propose effective tools to detect and mitigate the effects of cyber  threats;  see for example  \cite{sinapoli2014},  \cite{mishra2014secure}, \cite{sandberg2015cyberphysical},  \cite {teixeira2015secure:b}, \cite{zamaniautomatic2015} and the references listed therein.  Moreover, integration of  encryption techniques inside control and estimation paradigms, which only deal with encrypted data,  becomes an emerging research topic in recent years \cite{farokhi2016secure,kim2016encrypting,kogiso2015cyber}.

Consensus law is perhaps one of the most common protocols in neworked systems and has been widely studied not only by systems and control theorists, but also by physicists, biologists, sociologists, and mathematicians. In systems and control community, this protocol has been investigated under consensus of a multi-agent system (MAS) topic. The early works on consensus of MAS with  first-order dynamics can be found in  \cite{olfati-saber_consensus_2004,jadbabaie_coordination_2003}, in both continuous-time and	discrete-time settings.  Over the past decade, extensive efforts have been devoted to the research   
on more complicated systems including  second-order systems  \cite{ren_second-order_2007, xie_consensus_2012}, general linear systems \cite{wieland_2011,you_tac_2011},
and even nonlinear heterogeneous systems \cite{isidori_tac_2014,zhu_tac_2016}. 
It is known from the history of consensus theory that breakthrough from 
first-order dynamics to second-order ones was often significant in many scenarios;
see a survey paper \cite{knorn_cns_2016}.  It is worth mentioning that the consensus problem of 
a second-order MAS with time-varying topologies still remains open with many attempts in, e.g., \cite{Scardovi:09,meng:18},
while the same problem for a first-order MAS has been well studied many years ago in \cite{Moreau:05,Ren:05}.

 Given its wide range of applications, privacy and security challenges associated with consensus protocol is of great practical and theoretical importance. As consensus protocols require information exchange among agents within a network, if an agent's state is completely known by others at one time instant, its whole trajectory might be reconstructed
	using the knowledge of dynamics. In many practical situations, leakage of information is not allowed.
	Therefore, researchers are interested in preserving privacy of an agent, typically,  keeping  its initial states secret from others,
	during the evolution of network.
	A common privacy preserving approach is to use obfuscation to mask the true state values by
	adding random noises to the average consensus process; see, e.g.,  \cite{kefayati_secure_2007,mo_privacy_2017}.
	In this noise-based setting, differential privacy is another important tool for database privacy in
	computer science \cite{nozari_ifac_2015,huang_differentially_2012,manitara_privacy-preserving_2013}.
	The result for privacy preserving maximum consensus can be found in
	\cite{duan_privacy_2015} where all agents independently generate and transmit random numbers before sending out their initial states. The  noise-based obfuscation techniques  prevent
	forming consensus at the desired average value and achieve consensus in
	the statistical mean-square sense. To mitigate such effects on
	network performance, researchers are  proposing other secure
	control protocols.  Another common approach is to employ  tools from
	cryptography. 
	
Application of cryptography-based approaches to a 
	decentralized protocol like  consensus, especially without a trusted third-party, is  challenging.  An early result can be found in
	\cite{lagendijk_2013} where privacy is preserved at the cost of
	depriving participating  agents from access to the agreed value. A
	more complete result  has been reported in a recent paper
	\cite{ruan_secure_2017} for first-order  agents where
	a homomorphic  cryptography-based approach is used to guarantee
	privacy and security in a decentralized consensus scenario.
	Also, randomness is added to the coupling weights such that the
	transmission signals via network are of higher confidentiality.
		However, in their work, the privacy analysis result is only for first-order systems, where leakage of coupling weights directly leads to disclosure of agents' states. This phenomenon may not necessarily happen, depending on the network convergence rate,  in the case of agents with second-order dynamics.}

	In this paper, we attempt to propose a secure and privacy
	preserving  average consensus algorithm for second-order multi-agent systems, using a cryptography-based approach as well as
	random network coupling weights. The research for second-order multi-agent systems has
	two-fold contributions.

	On one hand, a protocol with random network coupling weights unavoidably leads to
	a network of time-varying weighted topologies. Therefore, the
	development relies on a consensus algorithm for time-varying topologies.
	In a discrete-time setting,  sufficient and necessary conditions for second-order multi-agent systems were revealed in \cite{xie_consensus_2012}, only for a fixed topology.
	The result in \cite{lin_automatica_2009} applies for time-varying topologies, but with a compulsory
	velocity damping, which results in agent velocities agreeing at zero.
	The scenarios  studied in \cite{su_automatica_2012} and \cite{huang_tac_2017} cover second-order systems, but only
	for neutrally stable systems whose eigenvalues  are semi-simple with modulus 1.
	The authors of \cite{yao_siam_2013} studied the general scenario as in this paper; however, with
	very restrictive network connectivity conditions including fully connected, neighbor shared, and at least quasi-neighbor shared networks.  A very recent paper for dealing with second-order systems with time-varying topologies
	can be found in \cite{meng_2018}, but only in a continuous-time setting.
	As the cornerstone for  establishing a secure and privacy preserving
	average consensus algorithm,  we first establish conditions for achieving consensus among second-order agents with time-varying topology and discrete time dynamics. This is the first contribution which can be regarded as extension of a result in \cite{xie_consensus_2012}.
	
%
The second contribution lies in development of an algorithm to preserve agents privacy within the network. The proposed cryptography-based consensus algorithm with random network weights guarantees privacy for an agent connected with more than one neighbors.  Moreover, we fully examine the scenarios under which leakage of private information might occur. It is worth  mentioning that in a network of second-order  agents, the two-dimensional agent states,
	say position and velocity, are
	encrypted and transmitted in a lumped quantity.  Hence, the secure and privacy preserving analysis
	becomes totally different from that of first-order systems. Moreover, it turns out that unlike the case of  first-order systems, for a second-order system scenario,
	the possibility of estimating a neighbor's initial state depends on the network convergence rate, even when the weights are fully known by agents.

	The rest of this paper is organized as follows. In Section~\ref{sec-Pre}, graph theory and Paillier encryption are briefly introduced. The main results including a consensus algorithm for
	second-order systems with time-varying topologies and its cryptography based version
	are proposed in Section~\ref{sec-MainResults}.
	The conditions for privacy preserving are also revealed in this section.
	Numerical examples are presented in Section~\ref{sec-Simlation}.
	Finally, some conclusions are drawn in Section~\ref{sec-Conclusion}.

	\section{Preliminaries}\label{sec-Pre}
	
	In this section, some preliminary concepts regarding graph theory and Paillier cryptosystem are briefly introduced.
	
	\subsection{Graph Theory}\label{ssec-GraphTheory}
	Let   $\mathcal{G}$($\mathbb{V}$,$\mathbb{E}$) denote a connected undirected graph. The set of all nodes and edges are presented by $\mathbb{V}$ and  $\mathbb{E}$, respectively.  The elements in  $\mathbb{V}$ are $\nu_1,\nu_2,\cdots,\nu_N$, where $N$ is the number of nodes, and the elements of the set $\mathbb{E}$ are pairs of nodes denoted by $(\nu_i,\nu_j)$.  In a network associated with graph $\mathcal{G}$($\mathbb{V}$,$\mathbb{E}$), each node $\nu_i$ represents an agent and each edge $(\nu_i,\nu_j)$ exists if the agents
	$i$ and $j$ communicate with each other.
	An adjacency matrix $\mathcal{A}$ is defined as follows. For every entry $a_{ij}$, i.e., coupling weight between $\nu_i$ and $\nu_j$, if $(\nu_j,\nu_i)\in \mathbb E$, there is $a_{ij}>0$, otherwise $a_{ij}=0$. And the Laplacian matrix is defined as $L=\{l_{ij}\}$, where
	$l_{ij}=-a_{ij},i\neq j$, and
	$l_{ii}=\sum_{j=1,j\neq i}^{N}a_{ij},$ $\forall \nu_i,\nu_j \in \mathbb V$.

	\subsection{Paillier Encryption }
	
	\label{ssec-Paillier}
	
	\subsubsection{Asymmetric Cryptosystem}
	
	In the asymmetric cryptosystem, there are two types of keys, namely, a private key $K^{Prv}$ available only to each individual for decryption and a public key $K^{Pub}$ accessible by all entities  only for encryption purposes.  Consider two individuals, say Alice and Bob, who intend to communicate in a secure way so that a third party, call it Eve, cannot recover the message. To achieve this, Alice generates a public key $K_A^{Pub}$ that everyone, perhaps including Eve, knows about. Then before Bob sends a message $m$ to Alice, it encrypts it as $\mathcal{E}_A(m)$  using $K_A^{Pub}$ provided by Alice.  However, the only way to decrypt the ciphertext $\mathcal{E}_A(m)$   is  through the private key $K_A^{Prv}$, which is only known by Alice.  Since Eve does not have  access to the private key $P^{Prv}_A$, information privacy remains intact.
	
	\subsubsection{Semi-homomorphic Property of Paillier Cryptosystem}
	
	The Paillier encryption  exploited in this paper is a certain class
	of asymmetric cryptosystem that adopts semi-homomorphic encryption
	techniques \cite{paillier_1999}.  Homomorphic encryption allows operations to be done
	on ciphertext such that after decryption the results remain
	equivalent to those obtained from performing some other operations
	on the associated plain text.
	A fully-homomorphic encryption has the property that for an arbitrary function $f$ operating on encrypted data $\mathcal{E}(m_i),i=1,2,\dots,n$,  which are obtained from encryption of messages $m_i$ under the same public key, there always exists a function $f'$ such that $\mathcal{E}(f'(m_1,m_2,\dots,m_n))=f(\mathcal{E}(m_1),\mathcal{E}(m_2),\dots,\mathcal{E}(m_n))$.
	In a semi-homomorphic cryptosystem,  the above property only holds for some limited classes of functions. For instance,   Paillier encryption has additive semi-homomorphic property. Under Paillier encryption, the following properties hold,
	\begin{equation}\label{equ-SemiHomo-Property}
	\begin{split}
	\mathcal{E}(m_1)\cdot\mathcal{E}(m_2)&=\mathcal{E}(m_1+m_2)\\
	\big(\mathcal{E}(m_1)\big)^n&=\mathcal{E}(n\cdot m_1),
	\end{split}
	\end{equation}
	where $m_1$ and $m_2$ are plaintexts and $\mathcal{E}$ denotes encryption function.
	
	A brief review of Paillier algorithm is demonstrated in Algorithm \ref{al:pailier}.
{\mz 	
	\begin{algorithm}
		\caption{Paillier Cryptosystem}\label{al:pailier}
	  \begin{algorithmic}
	  	 \State \textbf{Key Generation}:\\
		1. Generate 2 large prime number $p$, $q$, where\\ $\gcd(pq,(p-1)(q-1))=1$.\\
		2. $\lambda={\rm lcm}(p-1,q-1)$, and modulus $n=p\cdot q$.\\
		3. Define $Z_n=\{x|x\in Z, 0\le x<n\}$,\\
		$Z_n^*=\{x|x\in Z, 0\le x<n,\gcd(x,n)=1\}$.\\
		
		4. Select an integer $g\in  Z_{n^2}^* $ that satisfies
		\begin{equation*}
		\gcd(L(g^\lambda \mod n^2),n)=1,
		\end{equation*}
		where the function $L$ is defined by
		\begin{equation*}
		\begin{split}
		L(x)&=\frac{x-1}{n},\\
		x&\in \big\{x<n^2|x\equiv 1 \mod n \big\}.
		\end{split}
		\end{equation*}
		Public key is $(n,g)$. Private key is $(\lambda)$.\\
		
		\State \textbf{Encryption}:\\
		1. Generate a random integer $ r\in (0,n)$.\\
		2. Compute
		\begin{equation*}
		c=g^m\cdot r^n \mod n^2.
		\end{equation*}
		
		\State \textbf{Decryption}:\\
		1. Compute
		\begin{equation*}
					m=\frac{L(c^\lambda \mod n^2)}{L(g^\lambda \mod n^2)} \mod n
				\end{equation*}
		  \end{algorithmic}
		\end{algorithm}
	}

	\section{Main Results}\label{sec-MainResults}
	
	In this section, we study how to exploit Paillier cryptosystem to ensure information privacy of  agents with double integrator dynamics which are connected under the consensus law.  Our objective is to make sure that during the process of reaching consensus, the states of all agents remain secret from their neighbors. We present a novel strategy  which is capable of keeping agents' privacy over undirected graphs.  Moreover, we study conditions under which the information associated with an agent, i.e.,  its position or velocity, might be leaked to an eavesdropper or a curious neighbor. It is noted that the possibility of leakage  due to disclosure of private keys is out of the scope of the current paper and is not considered.

	\subsection{Second-order Systems}\label{ssec-2OrderProtocol}
	
	Consider a second-order system with its dynamics represented in discrete-time form as follows
	\begin{equation}\label{equ-2orderModel}
	\begin{split}
	{p}_i^{(k+1)}&=p_i^{(k)}+Tv_i^{(k)},\\
	{v}_i^{(k+1)}&=v_i^{(k)}+Tu_i^{(k)}, \quad i=1,2\dots,N,
	\end{split}
	\end{equation}
	where $k$ is the time index, $T$ is the sampling time, and $p_i, v_i\in \mathbf{R}$ are the position and velocity associated with the agent $i$,  respectively.   Without loss of generality, we let the sampling time  be unity throughout this paper.
	The following control law drives the group of agents  in \eqref{equ-2orderModel} toward reaching
	consensus asymptotically
	\begin{equation}\label{equ-2orderInput}
	u_i^{(k)}=\sum_{j\in \mathbb{N}_i}\gamma_1 a_{ij}^{(k)}(p_j^{(k)}-p_i^{(k)})
	+\gamma_2\sum_{j\in \mathbb{N}_i}a_{ij}^{(k)}(v_j^{(k)}-v_i^{(k)}),
	\end{equation}
	where $a_{ij}^{(k)}$s are entries of adjacency matrix $ \mathcal{A}$ derived from a connected and undirected network topology, $\gamma_1$, $\gamma_2$ are the coefficients and $\mathbb{N}_i$ is the set of agent $i$'s neighbors.
	
	Let
	\begin{equation*}
	\begin{split}
	p^{(k)}&=\left(p_1^{(k)},{p_2^{(k)}},\cdots,p_N{^{(k)}}\right)^\mathsf{T},\\
	v^{(k)}&=\left(v_1^{(k)},{v_2^{(k)}},\cdots,v_N{^{(k)}}\right)^\mathsf{T},\\
	x^{(k)}&=\left((p^{(k)})^\mathsf{T},({v^{(k)}})^\mathsf{T}\right)^\mathsf{T}.
	\end{split}
	\end{equation*}
	With the definition of the Laplacian matrix $L^{(k)}$,
	substituting  \eqref{equ-2orderInput} into \eqref{equ-2orderModel} yields
	\begin{equation}\label{equ-Systems}
	x^{(k+1)}=F^{(k)}x^{(k)},
	\end{equation}
	where  $F^{(k)}=
	\begin{pmatrix}
	I_N &  I_N\\
	-\gamma_1 L^{(k)} & I_N-\gamma_2 L^{(k)}
	\end{pmatrix}
	$.
	If the system has a fixed topology, the  dynamics in \eqref{equ-Systems} shrinks to
	\begin{equation}\label{equ-Systems-Fixed}
	x^{(k+1)}=Fx^{(k)}.
	\end{equation}

	For an undirected and connected topology, the Laplacian matrix $L$ attains an eigenvalue $0$ associated with a right eigenvector $\mathbf{1}_N$ and a left eigenvector $\mathbf{1}_N^{\mathsf{T}}$; see, e.g. \cite{godsil2013algebraic}.
	Thus, one can easily conclude that $F$ has eigenvalue $1$ associated with algebraic multiplicity two, associated with right eigenvectors $(\mathbf{1}_N^{\mathsf{T}},\mathbf{0}_N^{\mathsf{T}})^{\mathsf{T}}$ and $(\mathbf{0}_N^{\mathsf{T}},\mathbf{1}_N^{\mathsf{T}})^{\mathsf{T}}$. Moreover, since $L$ is symmetric,  the matrix $F$  has left eigenvectors $(\mathbf{1}_N^{\mathsf{T}},\mathbf{0}_N^{\mathsf{T}})$ and $(\mathbf{0}_N^{\mathsf{T}},\mathbf{1}_N^{\mathsf{T}})$.

	We need the following definitions which are required for the further developments presented in this paper.
	
	\begin{definition}\label{def-NormalConsensus}
		A group of discrete-time second-order agents with dynamics as in \eqref{equ-2orderModel} is said to reach (asymptotic) consensus if the following property holds for any initial conditions:
		\begin{equation*}
		\lim_{k\to+\infty}p_i^{(k)}-p_j^{(k)}=0,\quad \lim_{k\to+\infty}v_i^{(k)}-v_j^{(k)}=0,\quad
		\forall i,j \in \mathbb V.
		\end{equation*}
	\end{definition}
	
	The above definition is concerned about consensus as $k\to \infty$, which is an asymptotic property.
	To explicitly describe a consensus behavior in finite time, we introduce the following definition.
	
	\begin{definition}\label{def-PracticalConsensus}
		A group of discrete-time second-order agents with dynamics as in \eqref{equ-2orderModel} is said to reach $\delta$-practical consensus at time $k_c$ if
		\begin{equation}\label{equ-PracticalConsensusDefinition}
		|p_j^{(k)}-p_i^{(k)}|<\delta, \quad \forall k>k_c \text{ and } i,j \in \mathbb V.
		\end{equation}
	\end{definition}
	
	It is noted that the above definition demands for global knowledge of positions and/or velocities of agents. Therefore, we introduce the following definition that relies only on the local information.
	
	\begin{definition}\label{def-LocalAgreement}
		In a group of discrete-time second-order agents with dynamics as in  \eqref{equ-2orderModel}, two agents, say $i$ and $j$, are said to reach local $\delta$-agreement at time $k_a$ if
		\begin{equation}\label{equ-LocalAgreement}
		|p_j^{(k_a)}-p_i^{(k_a)}|\leq \delta.
		\end{equation}
		
	\end{definition}

	We now  revisit a lemma from \cite{xie_consensus_2012} that provides conditions for a group of agents with dynamics as in \eqref{equ-2orderModel} to reach consensus.
	
	\begin{lemma}\label{lem-Consensus-FixedTopology}
		
		Suppose that a group of discrete-time second-order agents with dynamics as in  \eqref{equ-2orderModel}
		is connected under a fixed topology with the control law \eqref{equ-2orderInput}.
		Then the closed-loop system in \eqref{equ-Systems-Fixed} reaches consensus with
		\begin{equation}\label{equ-Condition-UndirectedCosnensus}
		\begin{split}
		\left\{
		\begin{aligned}
		&\gamma_2>\gamma_1>0,\\
		&\gamma_1-2\gamma_2>\frac{-4}{\mu_i},
		\end{aligned}
		\right.
		\end{split}
		\end{equation}
		where $\mu_i$s are nonzero eigenvalues of the Laplacian matrix $L$,
		if and only if the associated topology is connected.
	\end{lemma}
	
	\noindent{\it Proof:} To be self-contained, the proof in \cite{xie_consensus_2012} is briefly presented here.
	Given the connectivity of the topology, the Laplacian matrix of the system has an eigenvalue $0$ associated with an eigenvector $\mathbf{1}_N$.
	The condition \eqref{equ-Condition-UndirectedCosnensus} ensures that the rest eigenvalues of $L$ are negative, which makes the remaining eigenvalues of $F$ stay within the unit circle.
	Let $J$ be the Jordan canonical form of $F$ associated with an invertible matrix $P$, then we have
	\begin{equation}\label{equ-Jordanform-F}
	\begin{split}
	F&=PJP^{-1}\\
	&=P
	\begin{bmatrix}
	\begin{matrix}
	1 & 1\\
	0 & 1
	\end{matrix} &
	\mathbf{0}
	_{2\times (2N-2)}
	\\  \mathbf{0}
	_{(2N-2)\times 2}
	& \tilde{J}_{(2N-2)\times(2N-2)}
	\end{bmatrix}
	P^{-1},\\
	\end{split}
	\end{equation}
	where $\tilde{J}$ is a Jordan canonical form matrix containing all the eigenvalues of $J$ except an eigenvalue $1$ with algebraic multiplicity two. The vectors $\frac{1}{N}(\mathbf{1}_N^{\mathsf{T}},\mathbf{0}_N^{\mathsf{T}})$ and $\frac{1}{N}(\mathbf{0}_N^{\mathsf{T}},\mathbf{1}_N^{\mathsf{T}})$ are the first and second rows of $P$, respectively.
	One should notice that all eigenvalues of $\tilde{J}$ are smaller than one and $\lim_{k\to\infty}\tilde{J}^k=\mathbf{0}$. Then by substituting \eqref{equ-Jordanform-F} into \eqref{equ-Systems-Fixed}, we calculate the norm
	\begin{equation}
	\begin{split}
	&\lim_{k\to\infty}
	\left\|
	\begin{bmatrix}
	p^{(k)}\\
	v^{(k)}
	\end{bmatrix}
	-
	\begin{bmatrix}
	\frac{1}{N}\sum_{j=1}^{N}(p_j^{(0)}+kv_j^{(0)})\\
	\frac{1}{N}\sum_{j=1}^{N}v_j^{(0)}
	\end{bmatrix}\otimes\mathbf{1}_N
	\right\|\\
	=&\lim_{k\to\infty}
	\left\|
	F^k
	\begin{bmatrix}
	p^{(0)}\\
	v^{(0)}
	\end{bmatrix}
	-
	\begin{bmatrix}
	\frac{1}{N}\sum_{j=1}^{N}(p_j^{(0)}+kv_j^{(0)})\\
	\frac{1}{N}\sum_{j=1}^{N}v_j^{(0)}
	\end{bmatrix}\otimes\mathbf{1}_N
	\right\|\\
	=&\mathbf{0}.
	\end{split}
	\end{equation}
	\eproof
	
	\begin{theorem}\label{thm-Stablity-ChangingWeight-Undirected}
		There exists  an admissible variation range
		$\delta_A>0$ such that  all agents in the aggregated model \eqref{equ-Systems} under  an undirected graph topology  reach consensus if the following conditions are satisfied
		\begin{equation}\label{cond-Laplacian-SmallChange}
		\begin{split}
		\left\{
		\begin{aligned}
		&\|  \mathcal{A}^{(k)}- \mathcal{A}^{(0)}\|<\delta_A, \;\forall k \geq 0, \\
		&\gamma_2>\gamma_1>0,\\
		&\gamma_1-2\gamma_2>\frac{-4}{\mu_i^{(0)}},
		\end{aligned}
		\right.
		\end{split}
		\end{equation}
		where $\mu_i^{(0)}$s are nonzero eigenvalues of the Laplacian matrix $L^{(0)}$,  $\mathcal{A}$ is the adjacency of topology matrix, and $\| \mathcal{A}\|$ denotes the max norm of $\mathcal{A}$.
	\end{theorem}

	\noindent{\it Proof:} Firstly we calculate the Jordan canonical form of  the matrix $F^{(0)}$:
	\begin{equation}\label{equ-JodanForm-F-Proof}
	\begin{split}
	F^{(0)}&=P J P^{-1}\\
	&=P
	\begin{bmatrix}
	\begin{matrix}
	1 & 1\\
	0 & 1
	\end{matrix} &
	\mathbf{0}
	_{2\times (2N-2)}
	\\  \mathbf{0}
	_{(2N-2)\times 2}
	& \tilde{J}_{(2N-2)\times(2N-2)}
	\end{bmatrix}
	P^{-1}.\\
	\end{split}
	\end{equation}
	Given that the matrices $F^{(k)}$ and $F^{(0)}$ share common eigenvectors, by using the same matrix $P$, we can decompose $F^{(k)}$ as
	\begin{equation}\label{equ-JodanForm-F(k)-Proof}
	\begin{split}
	F^{(k)}&=P H^{(k)} P^{-1}\\
	&=P
	\begin{bmatrix}
	\begin{matrix}
	1 & 1\\
	0 & 1
	\end{matrix} &
	\mathbf{0}
	_{2\times (2N-2)}
	\\  \mathbf{0}
	_{(2N-2)\times 2}
	& \tilde{H}^{(k)}_{(2N-2)\times(2N-2)}
	\end{bmatrix}
	P^{-1},\\
	\end{split}
	\end{equation}
	where $H^{(0)}=J$ and $\tilde{H}^{(0)}=\tilde{J}$.   One should note that using the same $P$, for $k=0,1,\dots$, makes $\tilde{H}^{(k)}$s not necessarily be Jordan blocks when $k\ge1$. However, given the structure of the matrices $F^{(k)}$s, all of them share two common left eigenvectors and two right eigenvectors, therefore $H^{(k)}$s have a common block
	$\begin{bmatrix}
	1 & 1\\
	0 & 1
	\end{bmatrix}$.
	
	We now apply the following  coordinate transformation
	\begin{equation}\label{equ-Coordinate-Transform}
	\bar{x}^{(k)}=P^{-1}x^{(k)}
	\end{equation}
	and let $\bar{x}^{(k)}=
	\begin{bmatrix}
	\sigma_k\\
	\zeta_k
	\end{bmatrix}$
	for $\sigma_k \in \mathbf{R}^{2}$  and $\zeta_k \in \mathbf{R}^{2N-2}$.
	One should note that consensus is reached if $\zeta_k\to \mathbf{0}$ as $k\to\infty$.
	
	Consider a Lyapunov function
	\begin{equation}\label{equ-Lyapunov}
	Y_k=\zeta_k^{\mathsf{T}} Q \zeta_k.
	\end{equation}
	With the help of \eqref{equ-Systems}, \eqref{equ-JodanForm-F(k)-Proof}, \eqref{equ-Coordinate-Transform} and  \eqref{equ-Lyapunov},  one can write
	\begin{equation}\label{equ-Lyapunov-Derivative}
	\begin{split}
	Y_{k+1}-Y_k&=\zeta_{k+1}^{\mathsf{T}} Q \zeta_{k+1} - \zeta_k^{\mathsf{T}} Q \zeta_k\\
	&=\zeta_k^{\mathsf{T}} ((\underbrace{\tilde{H}^{(k)})^{\mathsf{T}}Q\tilde{H}^{(k)}-Q}_{R^{(k)}}) \zeta_k.
	\end{split}
	\end{equation}
	According to Lemma \ref{lem-Consensus-FixedTopology}, the conditions \eqref{cond-Laplacian-SmallChange} ensure that all eigenvalues of $\tilde{H}^{(0)}$ are inside the unit circle, which means there exists a positive definite matrix $Q$ such that
	$R^{(0)}={((\tilde{J})}^{\mathsf{T}}Q\tilde{J}-Q)$ is negative definite. As the change in eigenvalues of a matrix is a continuous function of its entries, one can conclude that $R^{(k)}$  is also  a negative definite matrix for a slight variation in entries of the matrix $\mathcal A$.
	\eproof


	\subsection{Privacy-preserving Consensus In Undirected Networks}\label{ssec-UndirectedGraph}
	
  First of all,   preserving an agent's privacy throughout the paper is defined as  maintaining its initial states hidden from other agents.	In this subsection, we introduce a strategy for preserving privacy among  a group of agents with dynamics as in \eqref{equ-2orderModel} under an undirected network topology.
	In particular, we provide a paradigm that manages encryption and information exchange policies to maintain privacy of agents' information from their neighbors while achieving consensus among them.
	
	\subsubsection{Confidential Strategy}
	
	To explain our proposed method in this subsection, we focus on an agent, say Alice, which communicates with other agents within its neighbor set under an undirected topology. One should note that all agents are employing the same protocol as in \eqref{equ-2orderInput}, if there is no state information leakage happening in process of calculating $u_A^{(k)}$, i.e., Alice's control input, the privacy of all agents remains intact.
	
	The information exchange strategy is the same between Alice and its neighbors, thus without loss of generality, we only need to examine the interaction between Alice and one of its neighbors, say Bob.

	In the following, we examine the possibility of information leakage from Bob to Alice in the process of computation of $u_A^{(k)}$.
	To this end, let us expand Alice's command input as follows
	\begin{equation*}
	\begin{split}
	u_{A}^{(k)}&=\underbrace{\gamma_1 a_{AB}^{(k)}(p_B^{(k)}-p_A^{(k)})+\gamma_2 a_{AB}^{(k)}(v_B^{(k)}-v_A^{(k)})}_{u_{AB}^{(k)}}+\\ &
	\sum_{j\in\mathbb{N}_A-\{Bob\}}\gamma_1 a_{Aj}^{(k)}(p_j^{(k)}-p_A^{(k)})
	+\gamma_2 a_{Aj}^{(k)}(v_j^{(k)}-v_A^{(k)}),
	\end{split}
	\end{equation*}
	where $\mathbb{N}_A$ denotes set of Alice's neighbors and $a_{AB}^{(k)}$ is an entry of the adjacency matrix $\mathcal A$  corresponding to Alice and Bob. Then Bob's contribution in $u_A^{(k)}$ is
	\begin{equation}\label{def-ContributionOfBob}
	u_{AB}^{(k)}=\gamma_1 a_{AB}^{(k)}(p_B^{(k)}-p_A^{(k)})+\gamma_2 a_{AB}^{(k)}(v_B^{(k)}-v_A^{(k)}).
	\end{equation}
	Similarly, the contribution of Alice to $u_B^{(k)}$  can be written as $u_{BA}^{(k)}$.
	
	\begin{remark}\label{rmk-LeastAndNecessary-Ua}
		According to \eqref{equ-2orderInput} and  \eqref{def-ContributionOfBob}, $u_{AB}^{(k)}$ is necessary information that Alice requires from Bob to compute $u_A^{(k)}$, which is the input to update $v_A^{(k)}$. In this scenario, if the states of Bob can be deduced from $u_{AB}^{(k)}$,  its privacy cannot be guaranteed under any
	 secure  communication protocol.
	\end{remark}

	For a first-order system, $u_{AB}^{(k)}$ in \eqref{def-ContributionOfBob} reduces to
	\begin{equation*}
	u_{AB}^{(k)}= a_{AB}^{(k)}(p_B^{(k)}-p_A^{(k)})     \end{equation*}
	and the state of Bob can be directly computed from $u_{AB}^{(k)}$ with $a_{AB}^{(k)}$ known, i.e.,
	\begin{equation*}
	p_B^{(k)} =    \frac{u_{AB}^{(k)} }{a_{AB}^{(k)}}+p_A^{(k)}.   \end{equation*}
	The situation for a second-order system is more complicated
	as $u_{AB}^{(k)}$ in \eqref{def-ContributionOfBob} contains lumped quantity of
	position and velocity.  Nevertheless, the computation of Bob's states is still possible when it has  a sole neighbor. 
	More specifically, at the two steps of $k=0$ and $k=1$, the two messages received from Bob by Alice are 
	as follows:
	\begin{equation}\label{equ-TwoSteps_Input}
	\begin{split}
		u_{AB}^{(0)}=\gamma_1 a_{AB}^{(0)}(p_B^{(0)}-p_A^{(0)})+\gamma_2 a_{AB}^{(0)}(v_B^{(0)}-v_A^{(0)}),\\
		u_{AB}^{(1)}=\gamma_1 a_{AB}^{(1)}(p_B^{(1)}-p_A^{(1)})+\gamma_2 a_{AB}^{(1)}(v_B^{(1)}-v_A^{(1)}).
	\end{split}	
	\end{equation}
Also, Bob's states obey the second-order dynamics 
	\begin{equation}\label{equ-TwoSteps_Dynamic}
	\begin{split}
		p_B^{(1)}&=p_B^{(0)}+v_B^{(0)},\\
		v_B^{(1)}&=v_B^{(0)}+u_B^{(0)}.
	\end{split}
	\end{equation}
As the topology is undirected and Bob has  one sole neighbor, Alice is able to measure 
	\begin{equation}\label{equ-TwoSteps_OneNeighborUndirected}
		u_B^{(0)}= u_{BA}^{(0)} =-u_{AB}^{(0)}.
	\end{equation}
The  other four variables, i.e., $p_B^{(0)}, p_B^{(1)}, v_B^{(0)}$ and $v_B^{(1)}$,  can be computed from the set of four 
equations in (\ref{equ-TwoSteps_Input}) and (\ref{equ-TwoSteps_Dynamic}). In particular, Bob's initial states are
  	\begin{equation}\label{equ-TwoSteps-Results}
	\begin{split}
		p_B^{(0)}=&p_A^{(0)}-
		\frac{\gamma_1+\gamma_2}{\gamma_1^2}\frac{u_{AB}^{(0)}}{a_{AB}^{(0)}}+\frac{\gamma_2}{\gamma_1^2}\frac{u_{AB}^{(1)}}{a_{AB}^{(1)}}+\\
		&\frac{\gamma_2^2}{\gamma_1^2}(v_A^{(0)}-v_A^{(1)}-u_{AB}^{(0)}),\\
		v_B^{(0)}=&v_A^{(0)}-
		\frac{1}{\gamma_1}
		\left
		(\frac{u_{AB}^{(1)}}{a_{AB}^{(1)}}-\frac{u_{AB}^{(0)}}{a_{AB}^{(0)}}
		\right) +\\ &\frac{\gamma_2}{\gamma_1}(v_A^{(1)}-v_A^{(0)}+u_{AB}^{(0)})
		.
	\end{split}
	\end{equation}
Thus, Bob's privacy cannot be maintained irrelevant of which kind of security protocol is applied.

If Bob has more than one neighbors, the above computation becomes invalid. For this scenario,  
we establish privacy preserving conditions in the following theorem, 
for a group of agents with second-order dynamics under undirected topologies. 
Also, it is noted that the weights between agents and its neighbors are public information.
In order to analyze the privacy of agents' states during consensus process, let $\hat{p}_B^{(k)}$
	and $\hat{v}_B^{(k)}$ be the estimation values of Bob's position and velocity, calculated by Alice. Then the estimation errors of Bob's initial states are defined as follows.
	\begin{equation}\label{equ-DefineErrors}
	\begin{split}
	\varepsilon_p= \hat{p}_B^{(0)}-p_B^{(0)},\\
	\varepsilon_v = \hat{v}_B^{(0)}-v_B^{(0)}.
	\end{split}
	\end{equation}

	\begin{theorem}\label{thm-DisclosureWithoutDecouple}
		Consider two agents with dynamics as in \eqref{equ-2orderModel}, called Alice and Bob, which are connected  via an undirected network  topology under  the control law \eqref{equ-2orderInput}.
		Suppose $a_{AB}^{(k)}$ is public information.
		Then, when the local $\delta$-agreement between Alice and Bob is reached at time $k_a$, by collecting $u_{AB}^{(k)}$, $p_A^{(k)}$ and $v_A^{(k)}$, $k=1,2,\cdots,k_a$, Bob's initial position and velocity can be estimated by Alice with errors $\varepsilon_p$ and $\varepsilon_v$, respectively, which satisfy
		\begin{equation}\label{cond-EstimationError}
		\begin{split}
		|\varepsilon_p|\leq(\frac{\gamma_2}{\gamma_2-\gamma_1})^{k_a}\delta,\\
		|\varepsilon_v|\leq \frac{\gamma_1\gamma_2^{k_a-1}}{(\gamma_2-\gamma_1)^{k_a}}\delta.
		\end{split}
		\end{equation}
	\end{theorem}
	
	\noindent{\it Proof:}
	According to the control law \eqref{equ-2orderInput}, Alice obtains a set of equations based on the information $u_{AB}^{(k)}$ collected from $k=0$ to $k=k_a-1$, i.e.,
	\begin{equation}\label{equ-PrivacyInput}
	\begin{split}
	\left\{
	\begin{aligned}
	u_{AB}^{(0)}=&\gamma_1 a_{AB}^{(0)}(p_B^{(0)}-p_A^{(0)})
	+\gamma_2 a_{AB}^{(0)}(v_B^{(0)}-v_A^{(0)}), \\
	\vdots \\
	u_{AB}^{(k_a-1)}=&\gamma_1 a_{AB}^{(k_a-1)}(p_B^{(k_a-1)}-p_A^{(k_a-1)}) \\
	& +\gamma_2 a_{AB}^{(k_a-1)}(v_B^{(k_a-1)}-v_A^{(k_a-1)}).
	\end{aligned}
	\right.
	\end{split}
	\end{equation}
	Next, with the knowledge of \eqref{equ-2orderModel}, Alice can also  construct  another set of  equations as below
	\begin{equation}\label{equ-PrivacyDynamics}
	\begin{split}
	&\left\{
	\begin{aligned}
	p_B^{(1)}  = & p_B^{(0)}+v_B^{(0)}, \\
	&\vdots \\
	p_B^{(k_a)}  = & p_B^{(k_a-1)}+v_B^{(k_a-1)}.
	\end{aligned}
	\right.
	\end{split}
	\end{equation}
	At time $k_a$,  define
	\begin{equation}\label {equ-hat-p_B}
	\delta_p =p_B^{(k_a)} - \hat{p}_B^{(k_a)}.
	\end{equation}
	By setting 
	$\hat{p}_B^{(k_a)}=p_A^{(k_a)}$, one has $|\delta_p|\leq \delta$
	by Definition \ref{def-LocalAgreement}.
	
	Substituting \eqref{equ-hat-p_B} into  the last equation of \eqref{equ-PrivacyDynamics} provides
	\begin{equation}\label{equ-PrivacyDynamics-hat}
	\begin{split}
	&\left\{
	\begin{aligned}
	p_B^{(1)}  = & p_B^{(0)}+v_B^{(0)},\\
	&\vdots \\
	p_B^{(k_a-1)} = & p_B^{(k_a-2)} + v_B^{(k_a-2)},\\
	\hat{p}_B^{(k_a)} = & p_B^{(k_a-1)} + v_B^{(k_a-1)}-\delta_p.\\
	\end{aligned}
	\right.\\
	\end{split}
	\end{equation}
	Next,    by eliminating  $v^{(k)}_B, k=0,\ldots, k_a-1,$  in \eqref{equ-PrivacyInput} using  \eqref{equ-PrivacyDynamics},  one obtains the following set of equations, for $k=0,\ldots,k_a-1$,
	\begin{equation}\label{equ-Eliminate_Velocity}
	(\gamma_2-\gamma_1)p_B^{(k)}=\gamma_2 p_B^{(k+1)}-\gamma_1 p_A^{(k)}-\gamma_2 v_A^{(k)}-\frac{u_{AB}^{(k)}}{a_{AB}^{(k)}}.
	\end{equation}
	Given the above equation by invoking \eqref{equ-hat-p_B}, we can express the position and velocity of Bob as
	\begin{equation}\label{equ-DisclosureOfTrajectoryWithoutDecouple}
	\begin{split}
	p_B^{(k)}=&(\frac{\gamma_2}{\gamma_2-\gamma_1})^{k_a-k}\hat{p}_B^{(k_a)}
	+\sum_{T=0}^{k_a-k-1}(\frac{\gamma_2}{\gamma_2-\gamma_1})^T\varphi_{k+T} \\ &
	+(\frac{\gamma_2}{\gamma_2-\gamma_1})^{k_a-k}\delta_p,\\
	v_B^{(k)}=&-\frac{\gamma_1}{\gamma_2}(\frac{\gamma_2}{\gamma_2-\gamma_1})^{k_a-k}\hat{p}_B^{(k_a)}
	-\frac{\gamma_2-\gamma_1}{\gamma_2}\varphi_k
	\\ & -\frac{\gamma_1}{\gamma_2}\sum_{T=0}^{k_a-k-1}(\frac{\gamma_2}{\gamma_2-\gamma_1})^T\varphi_{k+T}  -\frac{\gamma_1}{\gamma_2}(\frac{\gamma_2}{\gamma_2-\gamma_1})^{k_a-k}\delta_p,\\
	k&=0,1,\dots,k_a,
	\end{split}
	\end{equation}
	where
	\begin{equation*}
	\varphi_k=\frac{1}{\gamma_1-\gamma_2}(\gamma_1 p_A^{(k)}+\gamma_2 v_A^{(k)}+\frac{u_{AB}^{(k)}}{a_{AB}^{(k)}}).
	\end{equation*}
	
	By letting $k=0$ in \eqref{equ-DisclosureOfTrajectoryWithoutDecouple},   Bob's initial states, i.e.,  its position and velocity, can be obtained as
	\begin{equation}\label{equ-InitialStatesBob}
	\begin{split}
	p_B^{(0)}=&(\frac{\gamma_2}{\gamma_2-\gamma_1})^{k_a} \hat{p}_B^{(k_a)}+
	\sum_{k=0}^{k_a-1}(\frac{\gamma_2}{\gamma_2-\gamma_1})^{k}\varphi_k \\
	&+(\frac{\gamma_2}{\gamma_2-\gamma_1})^{k_a}\delta_p,\\
	v_B^{(0)}=&-\frac{\gamma_1}{\gamma_2}(\frac{\gamma_2}{\gamma_2-\gamma_1})^{k_a}\hat{p}_B^{(k_a)}
	-\frac{\gamma_2-\gamma_1}{\gamma_2}\varphi_0 \\
	& -\frac{\gamma_1}{\gamma_2}\sum_{k=0}^{k_a-1}(\frac{\gamma_2}{\gamma_2-\gamma_1})^k\varphi_{k}-\frac{\gamma_1}{\gamma_2}(\frac{\gamma_2}{\gamma_2-\gamma_1})^{k_a}\delta_p.
	\end{split}
	\end{equation}
	
	Alice can estimates Bob's initial states, i.e., $p_B^{(0)}$ and $v_B^{(0)}$ through the following estimation law : 
	\begin{equation}\label{equ-InitialStatesBob-Estimation}
	\begin{split}
	\hat{p}_B^{(0)}=&(\frac{\gamma_2}{\gamma_2-\gamma_1})^{k_a} \hat{p}_B^{(k_a)}+
	\sum_{k=0}^{k_a-1}(\frac{\gamma_2}{\gamma_2-\gamma_1})^{k}\varphi_k,\\
	\hat{v}_B^{(0)}=&-\frac{\gamma_1}{\gamma_2}(\frac{\gamma_2}{\gamma_2-\gamma_1})^{k_a}\hat{p}_B^{(k_a)}
	-\frac{\gamma_2-\gamma_1}{\gamma_2}\varphi_0 \\
	& -\frac{\gamma_1}{\gamma_2}\sum_{k=0}^{k_a-1}(\frac{\gamma_2}{\gamma_2-\gamma_1})^k\varphi_{k}.
	\end{split}
	\end{equation}
	
	Finally, with respect to \eqref{equ-DefineErrors}, \eqref{equ-InitialStatesBob} and \eqref{equ-InitialStatesBob-Estimation}, the estimation error of  Bob's states  computed by Alice are  as follows,
	\begin{equation}\label{equ-Proof_Laststep}
	\begin{split}
	\varepsilon_p=&(\frac{\gamma_2}{\gamma_2-\gamma_1})^{k_a}\delta_p,\\
	\varepsilon_v=&\frac{\gamma_1}{\gamma_2}(\frac{\gamma_2}{\gamma_2-\gamma_1})^{k_a}\delta_p.
	\end{split}
	\end{equation}
	Since $|\delta_p|<\delta$, \eqref{equ-Proof_Laststep} implies  \eqref{cond-EstimationError} and the  proof is finished.
	\eproof
	
	\begin{remark} For the sequence $k=0, 1,\cdots$, one can define
		\begin{equation*}
		\delta(k) = |p_A^{(k)} - P_B^{(k)}|.
		\end{equation*}
		Obviously, Alice and Bob reaches local $\delta(k)$-agreement at time $k$.
		Theorem~\ref{thm-DisclosureWithoutDecouple} claims that the estimation error
		of Bob's initial position at time $k$ is bounded by
		\begin{equation*}
		|\varepsilon_p(k)| \leq (\frac{\gamma_2}{\gamma_2-\gamma_1})^{k}\delta(k)
		=\frac {\delta(k)}{(\frac{\gamma_2-\gamma_1}{\gamma_2})^{k}}.
		\end{equation*}
		In a consensus process, one has $\delta(k) \rightarrow 0$ as $k\rightarrow\infty$.
		If the consensus convergence is sufficiently fast in the sense of
		\begin{equation*}
		\frac {\delta(k)}{(\frac{\gamma_2-\gamma_1}{\gamma_2})^{k}}\rightarrow 0,
		\; as \; k\rightarrow\infty,
		\end{equation*}
		Alice may estimate Bob's initial position with a sufficiently small
		estimation error. However, if the consensus convergence is not fast,
		sufficiently precise estimation of Bob's initial position becomes impossible,
		that is, Bob's privacy is preserved.  The same arguments also hold for estimation
		of Bob's initial velocity.
	\end{remark}
	
	\begin{remark}
		It is worthwhile noting that  the computation of estimation error in \eqref{cond-EstimationError}  depends on achieving local $\delta$-agreement at  time $k_a$. However, in a privacy-preserving consensus states of neighbors are secret information, which means $k_a$  cannot be obtained   directly by agents. Consider the two agents   with specifications stated in Theorem \ref{thm-DisclosureWithoutDecouple}.  Alice has only access to its own states and Bob's message $u_{AB}^{(k)}$.   Since the states of Bob remain hidden from Alice, it becomes impossible for it to directly attain information of $k_a$ even with the knowledge of  permissible error $\delta$.  Furthermore,  in a second-order system, $u_{AB}^{(k)}=0$ does not imply $|p_B^{(k)}-p_A^{(k)}|=0$ or $|v_B^{(k)}-v_A^{(k)}|=0$. However, it is possible to indirectly estimate the value of $k_a$ using the $u_{AB}^{(k)}.$
		
	\end{remark}

	\subsubsection{Exchanging Information And Operation}\label{sssec-ExchangeMethod}
	
	Theorem~\ref{thm-DisclosureWithoutDecouple}  demonstrates that in an undirected topology the knowledge of link weights,  i.e., $a^{(k)}_{ij}$, relating two neighbors, say Alice and Bob, enable Alice to reconstruct the states of Bob by exploiting the collected inputs $u_{AB}^{(k)}$ from Bob and its own states, provided that the consensus convergence rate is sufficiently
	large. To overcome this shortcoming, we exploit a methodology initially introduced by \cite{ruan_secure_2017} for networks of agents with first-order dynamics. In this approach,  the weight $a_{AB}^{(k)}$ is hidden from both Alice and Bob by being decoupled into two factors, with one factor stored by Alice, called $a^{(k)}_A$, and the other by Bob, called $a^{(k)}_B$.
	In this technique, the weighting between Alice and Bob can be written as
	\begin{equation}\label{equ-UndirectedWeightDecouple}
	a_{AB}^{(k)}=a_A^{(k)}\cdot a_B^{(k)}.
	\end{equation}
	The decoupled weights $a_A^{(k)}$ and $a_B^{(k)}$ are  generated  randomly in a specified range so that the first inequality in \eqref{cond-Laplacian-SmallChange}  is satisfied. More specifically, one can assign uniform samples for $a_A^{(k)}$  and $a_B^{(k)}$ from the range $(\sqrt{a_{AB}^{(0)}-\delta_A},\sqrt{a_{AB}^{(0)}+\delta_A})$.
	

	We introduce Algorithm~\ref{alg-UnidrectedGraph}  that exploits the Paillier cryptosystem to maintain the privacy of information between Alice and Bob. In Algorithm \ref{alg-UnidrectedGraph}, for the sake of simplicity of notation we omit the parameter $k$.

	\begin{algorithm}
		\caption{Information exchange in undirected networks}
		\label{alg-UnidrectedGraph}
		
		\textbf{Preparation} (Alice):\\
		(1) At initial time $k=0$, generate a pair of public key $K_A^{Pub}$ and private key $K_A^{Prv}$, then send $K_A^{Pub}$ to all its neighbors, including Bob. \\
		(2) At time $k$, generate a random number  $a_{A}\in(\sqrt{a_{AB}^{(0)}-\delta_A},\sqrt{a_{AB}^{(0)}+\delta_A})$.\\
		\textbf{Preparation} (Bob):\\
		(1) At time $k$, generate a random number   $a_{B}\in(\sqrt{a_{AB}^{(0)}-\delta_A},\sqrt{a_{AB}^{(0)}+\delta_A})$.\\
		
		\textbf{Step 1} (Alice):\\
		(1.1) Encrypt position: $p_A\rightarrow-p_A\xrightarrow{\mathcal E}\mathcal{E}_A(-p_A)\xrightarrow{sent\;to}$ Bob.
		\\
		(1.2) Encrypt velocity: $v_A\rightarrow-v_A\xrightarrow{\mathcal E}\mathcal{E}_A(-v_A)\xrightarrow{sent\;to}$ Bob.
		
		\textbf{Step 2} (Bob):\\
		(2.1) Operate position: $p_B\xrightarrow{\mathcal E}\mathcal{E}_A(p_B)\rightarrow\mathcal{E}_A(p_B)\cdot\mathcal{E}_A(-p_A)=\mathcal{E}_A(p_B-p_A)\rightarrow(\mathcal{E}_A(p_B-p_A))^{\gamma_1 a_B}=\mathcal{E}_A(\gamma_1 a_B\cdot(p_B-p_A))$.
		\\
		(2.2) Operate velocity: $v_B\xrightarrow{\mathcal E}\mathcal{E}_A(v_B)\rightarrow\mathcal{E}_A(v_B)\cdot\mathcal{E}_A(-v_A)=\mathcal{E}_A(v_B-v_A)\rightarrow(\mathcal{E}_A(v_B-v_A))^{\gamma_2 a_B}=\mathcal{E}_A(\gamma_2 a_B\cdot(v_B-v_A))$.
		\\
		(2.3) Combine $p$ and $v$:
		$\mathcal{E}_A(\gamma_1 a_B\cdot(p_B-p_A))\cdot\mathcal{E}_A(\gamma_2 a_B\cdot(v_B-v_A))=\mathcal{E}_A(\gamma_1 a_B\cdot(p_B-p_A)+\gamma_2 a_B\cdot(v_B-v_A))\xrightarrow{sent\;to}$ Alice.
		
		\textbf{Step 3} (Alice):
		\\
		Decrypt and operate:\\
		$\mathcal{E}_A(\gamma_1 a_B\cdot(p_B-p_A)+\gamma_2 a_B\cdot(v_B-v_A))\xrightarrow{\mathcal{E}^{-1}} \gamma_1 a_B\cdot(p_B-p_A)+\gamma_2 a_B\cdot(v_B-v_A)\xrightarrow{\times a_A}\gamma_1 a_{AB}\cdot(p_B-p_A)+\gamma_2 a_{AB}\cdot(v_B-v_A)=u_{AB}$.
	\end{algorithm}
	
	\begin{theorem}\label{thm-Confidential.Undirected}
		Consider a group of agents  with  second-order  dynamics as in \eqref{equ-2orderModel} under the control law \eqref{equ-2orderInput} and the conditions \eqref{cond-Laplacian-SmallChange}.  Given that all agents adopt Algorithm~\ref{alg-UnidrectedGraph}  asymptotic consensus can be reached. Moreover, if an agent has  more than one neighbor its initial states cannot be learned by them.      \end{theorem}
	
	\indent{\it Proof:}
	Using the semi-homomorphic property of Paillier encryption and Theorem \ref{thm-Stablity-ChangingWeight-Undirected},   asymptotic consensus can be easily established. If Alice is  Bob's solo neighbor, states of Bob will be available to Alice after consensus is reached at some time say, $k_c$. This is because in this case we have  $u_{AB}^{(k)}=u_A^{(k)}$ and when consensus is reached $v_A^{(k_c)}=v_B^{(k_c)}$.
	Moreover, according to \eqref{equ-2orderInput}, in  an undirected network  it holds that $u_{AB}^{(k)}+ u_{BA}^{(k)}=0$.  Then Alice can calculate Bob's state $v_B^{(k)}$ using
	\begin{equation}
	v_B^{(k)}=v_A^{(k_c)}+\sum_{i=k}^{k_c-1}u_{AB}^{(i)}
	\end{equation}
	This holds irrespective of the encryption method Bob and Alice exploiting in their communication.    Now if Bob has more than one neighbors,  since the decoupled term  $a_B^{(k)}$ is only kept by Bob, substituting $a_{AB}^{(k)}$  into equation \eqref{equ-PrivacyInput} with $a_A^{(k)}\cdot a_B^{(k)}$ provides Alice more unknowns but no additional equations. This makes the set of  equations collected by Alice be unsolvable. Therefore privacy of Bob's states remains intact. This completes the proof.
	\eproof
	
 Theorems~\ref{thm-DisclosureWithoutDecouple} and \ref{thm-Confidential.Undirected}
provide the conditions for privacy preserving consensus of second-order systems, in the  cases  of known coupling weights and unknown coupling weights, respectively.
The corresponding conditions for first-order systems have been studied in \cite{ruan_secure_2017}. 
Table~\ref{tab-disscusion} outlines different scenarios that might raise depending on the agents' dynamics, their number of neighbors and availability of coupling weights information. It demonstrates the possible violation of  individual's  privacy and time required for that to happen.

\begin{table}[ht] 		
\centering
\begin{tabular}{|p{1cm}|p{1.6cm}|p{1.8cm}|p{1.8cm}|}
\hline
 & \multicolumn{3}{c|}{Coupling Weights} \\
\hline
No. of Neighbors & known (1st-order)  & known $ \; $ (2nd-order) & unknown \\
\hline\hline
1 &  $1$ step & $2$ steps & when consensus reached\\
\hline
$\geq 2$ & $1$ step &  depending on consensus convergance speed & never \\
\hline\end{tabular}
\captionsetup{justification=centering}
\caption{
		Possibility of privacy violation for an agent and its required time. 
	}
\label{tab-disscusion}
\end{table}


	\section{Simulation}\label{sec-Simlation}
	
	In this section, some numerical examples are  presented to demonstrate how the privacy is preserved under  Algorithm \ref{alg-UnidrectedGraph}.
	We implement the Paillier method written in C language \cite{Paillier}. During the simulation, modular bits used for generating keys of Paillier encryption are set to 64.
	
\begin{figure}[t]
		\centering
		\includegraphics[scale=.5]{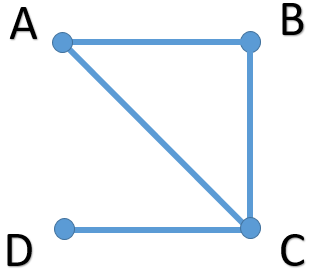}
		\caption{An undirected network composed of four nodes.}\label{fig-Topology}
	\end{figure}

	We consider the
	topology  in Fig. \ref{fig-Topology} with the associated Laplacian matrix
	$0.1 \times \begin{bmatrix}
	2 & -1& -1 & 0\\
	-1 & 2 & -1 & 0\\
	-1 & -1 & 3 & -1\\
	0 & 0 & -1 & 1
	\end{bmatrix}$.  Moreover, the other parameters  are set  as $\gamma_1=0.3$, $\gamma_2=0.6$, $p_A^{(0)}=20$, $p_B^{(0)}=30$, $p_C^{(0)}=50$, $p_D^{(0)}=90$, $v_A^{(0)}=30$, $v_B^{(0)}=-20$, $v_C^{(0)}=10$, and  $v_D^{(0)}=-40$.

	Since the communication protocols between every two agents are same, we focus on the states of  agents A and B in Fig. \ref{fig-Topology}.
	
\begin{figure}[tb]
		\begin{minipage}{0.45\textwidth}
			\centering
			\includegraphics[scale=.45]{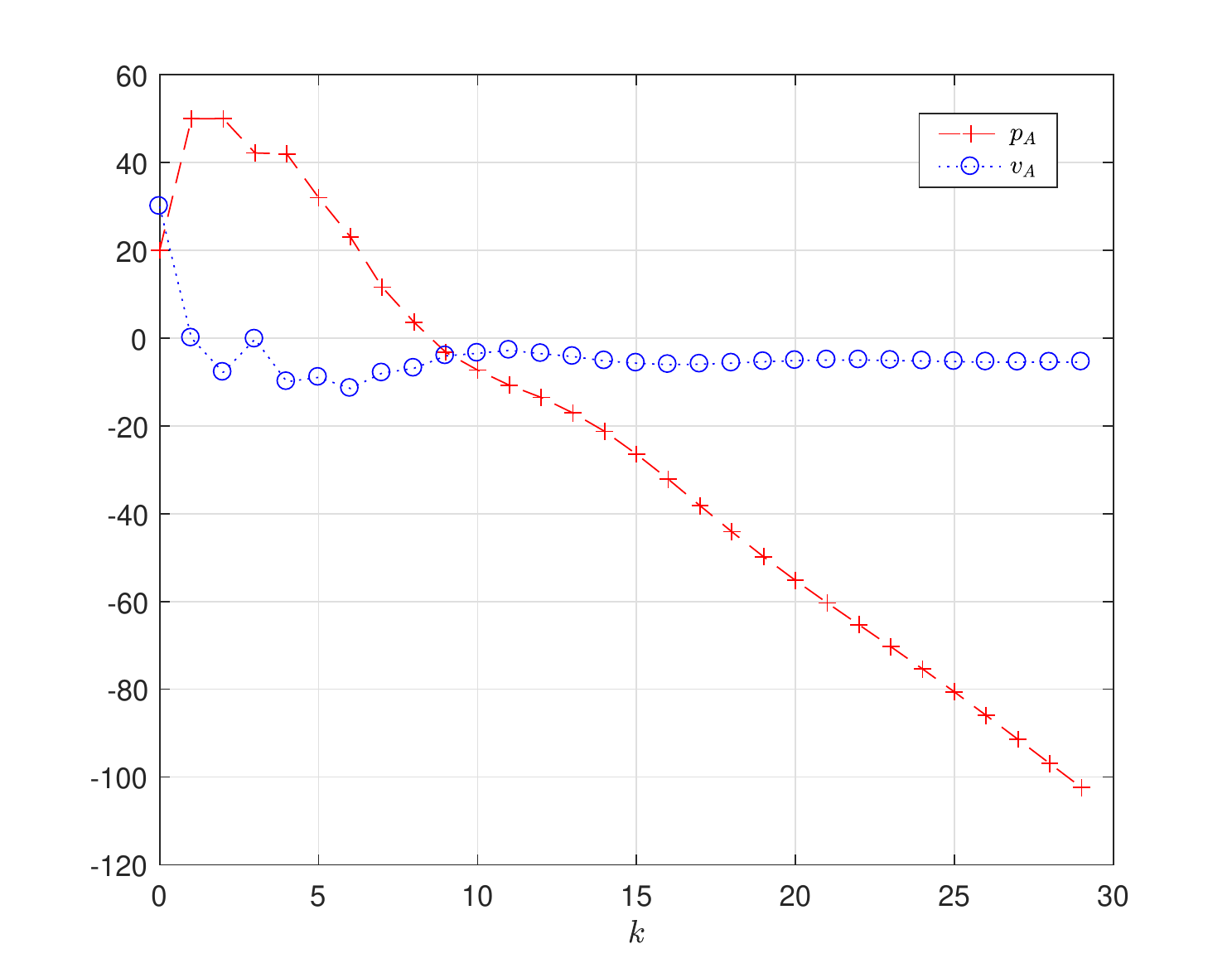}
			\subcaption{States of  agent A.}
		\end{minipage}
		\begin{minipage}{0.45\textwidth}
			\centering
			\includegraphics[scale=.45]{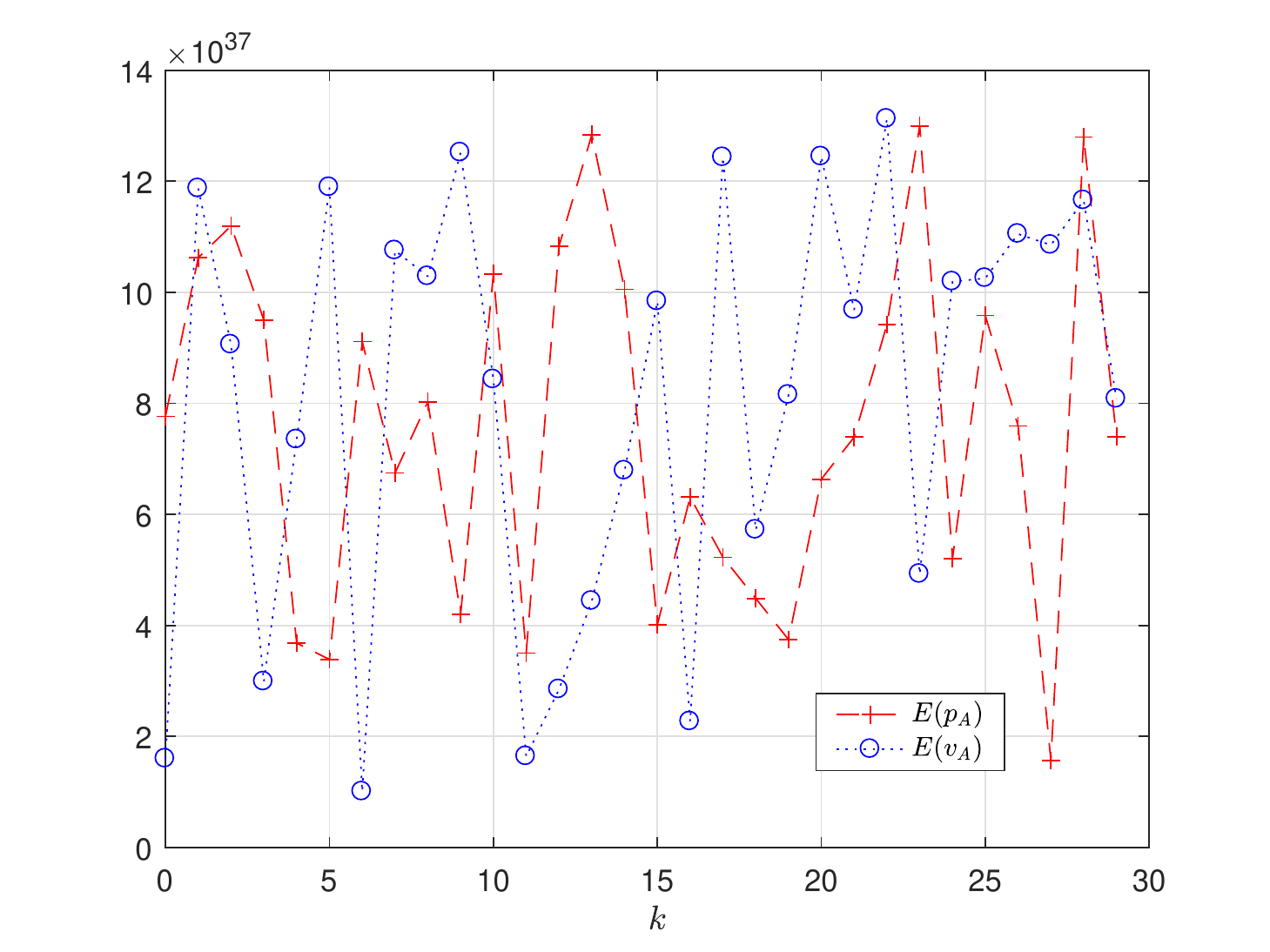}
			\subcaption{Encrypted states of   agent A as in step 1.1 and 1.2 of Algorithm~\ref{alg-UnidrectedGraph}.}
		\end{minipage}
		\begin{minipage}{0.45\textwidth}
			\centering
			\includegraphics[scale=.45]{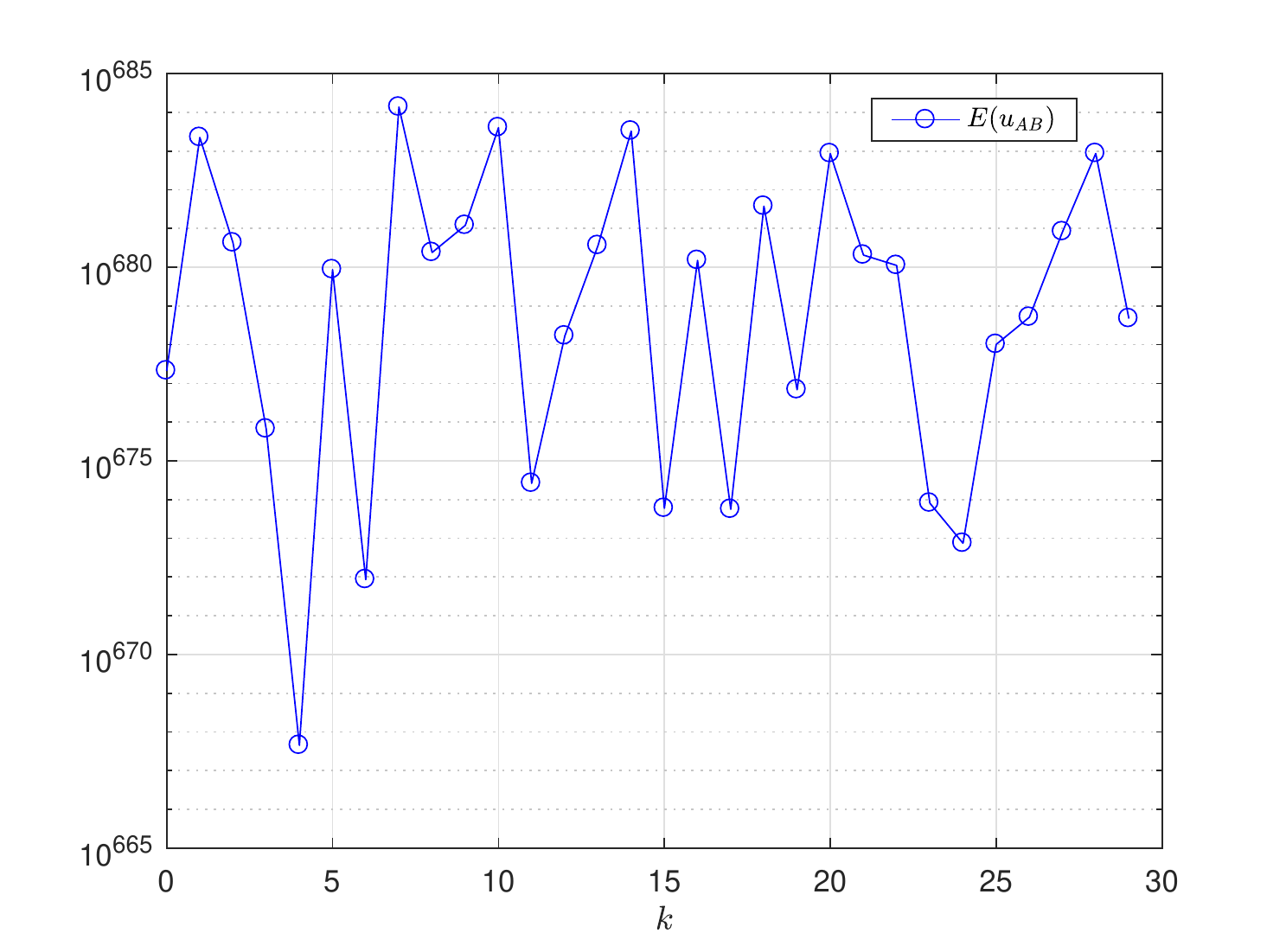}
			\subcaption{Message sent from agent B to A  as in Step 2.3 of Algorithm~\ref{alg-UnidrectedGraph}. }
		\end{minipage}
		\begin{minipage}{0.45\textwidth}
			\centering
			\includegraphics[scale=.45]{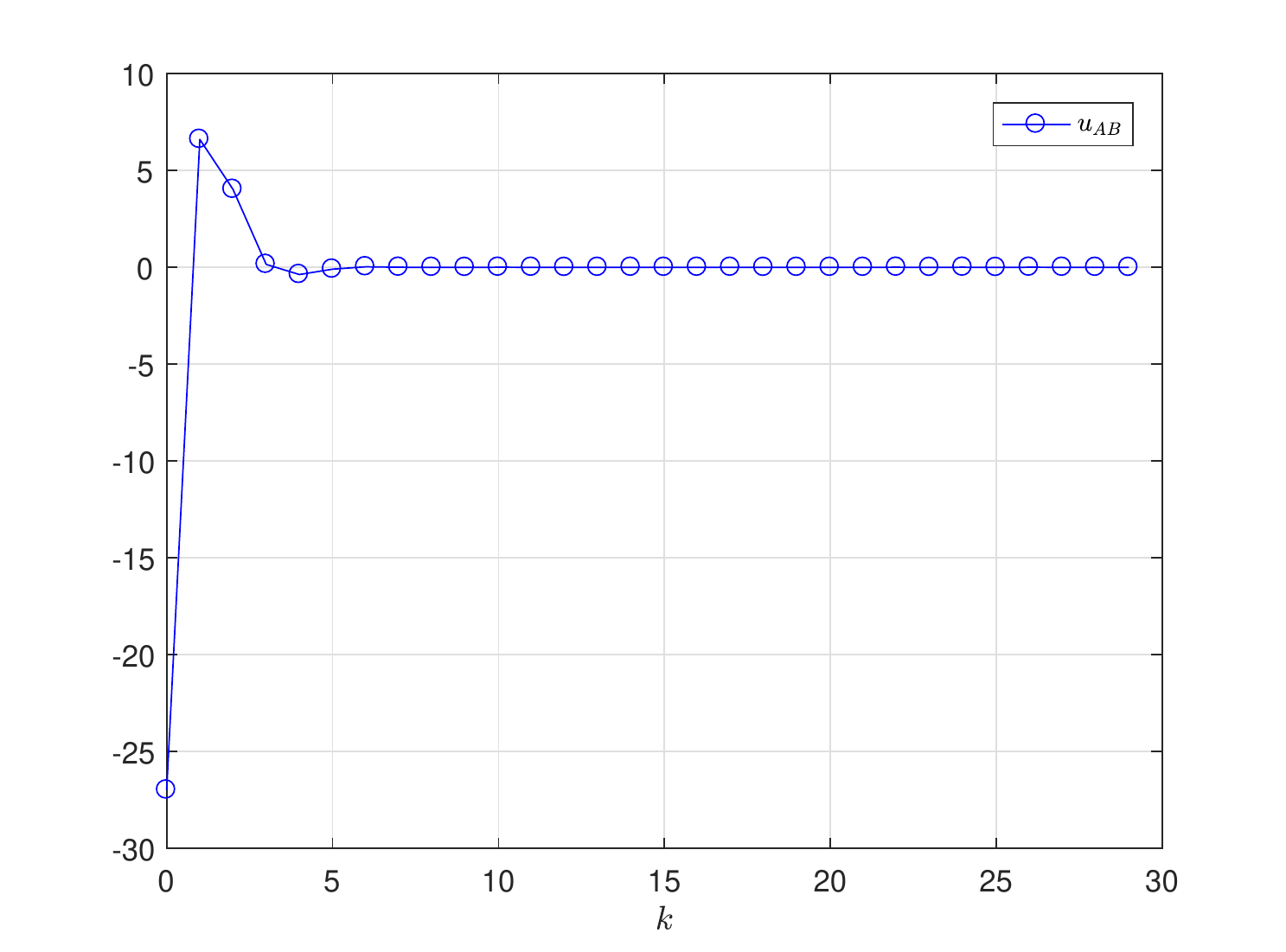}
			\subcaption{Decrypted version of the message sent from B to A as in step 3 of Algorithm~\ref{alg-UnidrectedGraph}.}
		\end{minipage}
		\caption{Exchanged information under Algorithm~\ref{alg-UnidrectedGraph}. }
		\label{fig-Algorithm_Process}
	\end{figure}

	Fig.~\ref{fig-Algorithm_Process} demonstrates how the information is exchanged under Algorithm~\ref{alg-UnidrectedGraph}.   In this figure, we present states, i.e., position and velocity of agent A, as well as information sent from agent B to A, i.e., $u_{AB}^{(k)}$. {\zc In Algorithm~\ref{alg-UnidrectedGraph}, if a malicious party  hacks into the communication channel  between A and B, the available information is $E_A(-p_A^{(k)})$, $E_A(-v_A^{(k)})$ and $u_{AB}^{(k)}$, which is transmitted in steps $1$-$3$, respectively. The figures show that the encrypted messages
contain no useful information for the malicious party and thus ensure network security. 	}
	
\begin{figure}[htb]
		\begin{minipage}{0.45\textwidth}
			\centering
			\includegraphics[scale=.45]{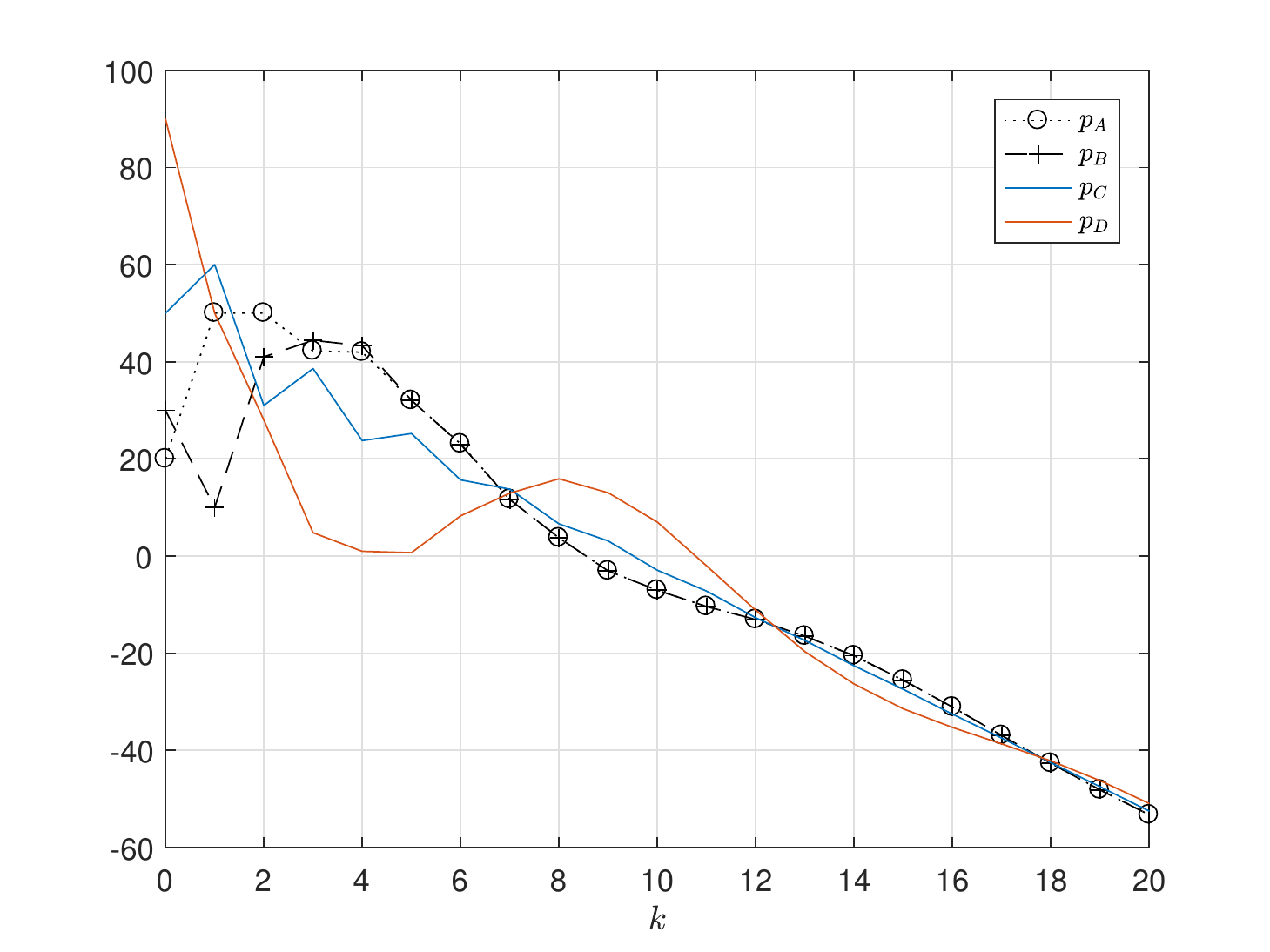}
			\subcaption{Position trajectories of all four agents.}
		\end{minipage}
		\begin{minipage}{0.45\textwidth}
			\centering
			\includegraphics[scale=.45]{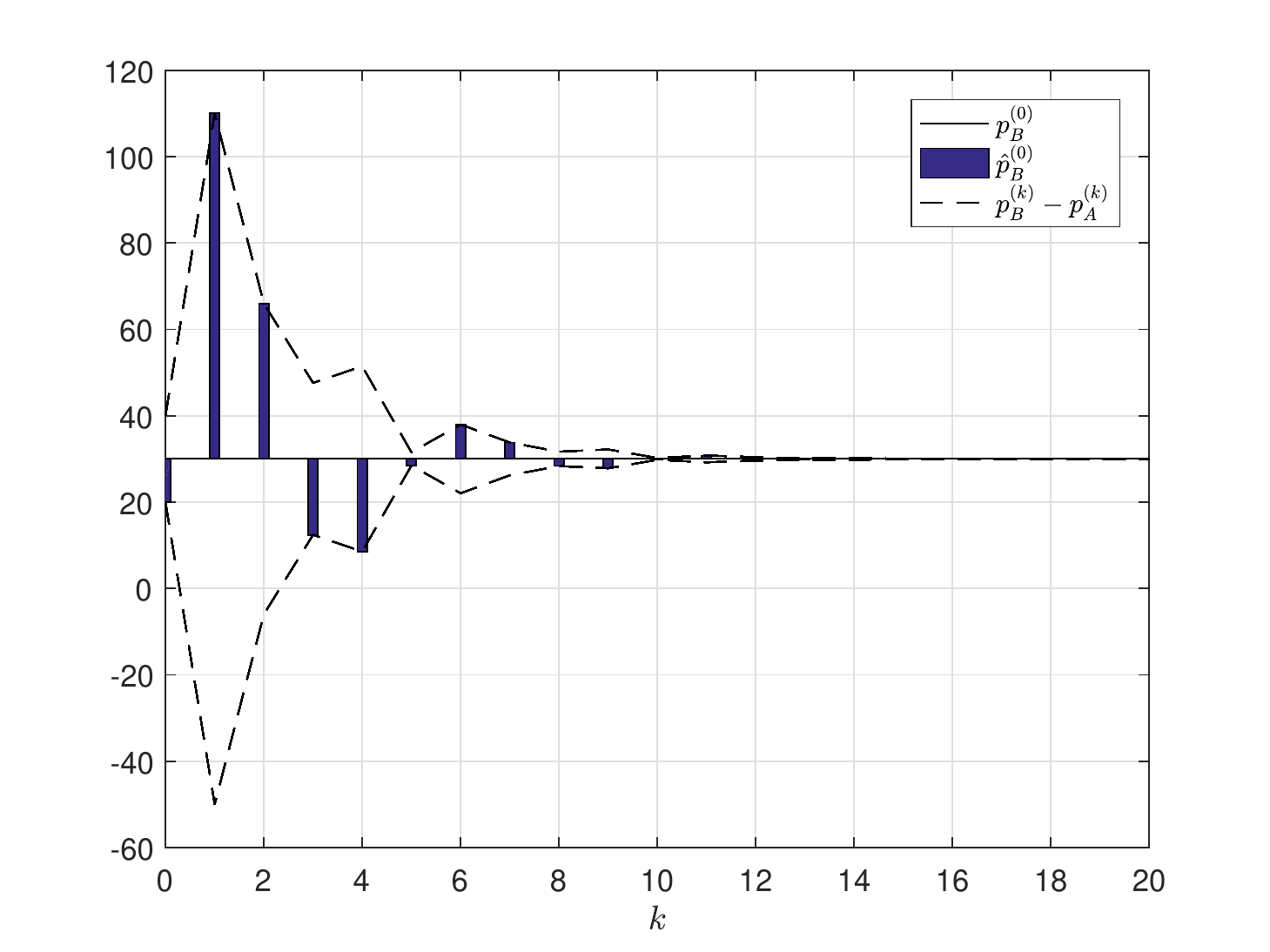}
			\subcaption{The initial position of agent B and its associated estimates computed by agent A.}
		\end{minipage}
		\begin{minipage}{0.45\textwidth}
			\centering
			\includegraphics[scale=.45]{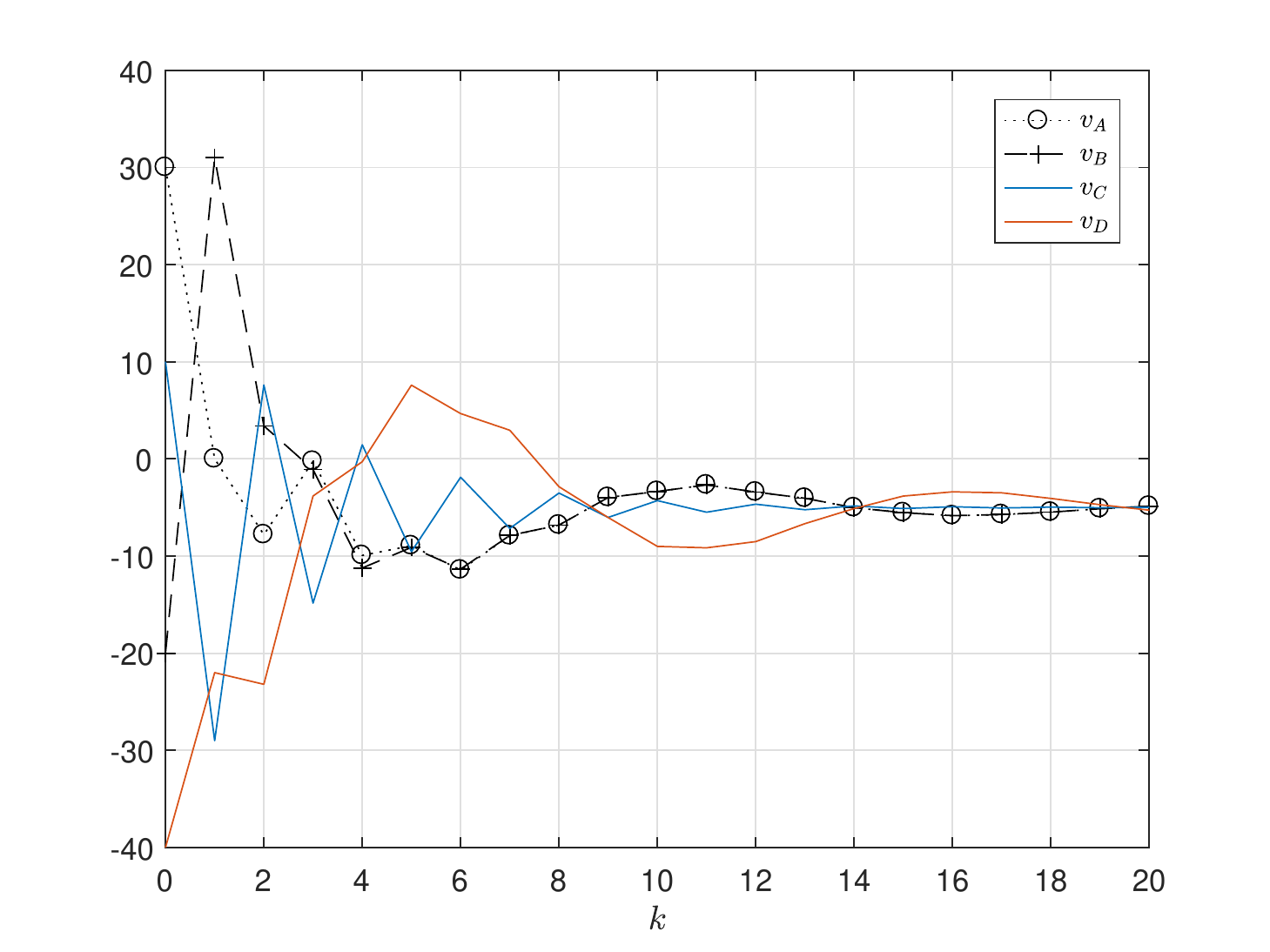}
			\subcaption{Velocity trajectories of all four agents.}
		\end{minipage}
		\begin{minipage}{0.45\textwidth}
			\centering
			\includegraphics[scale=.45]{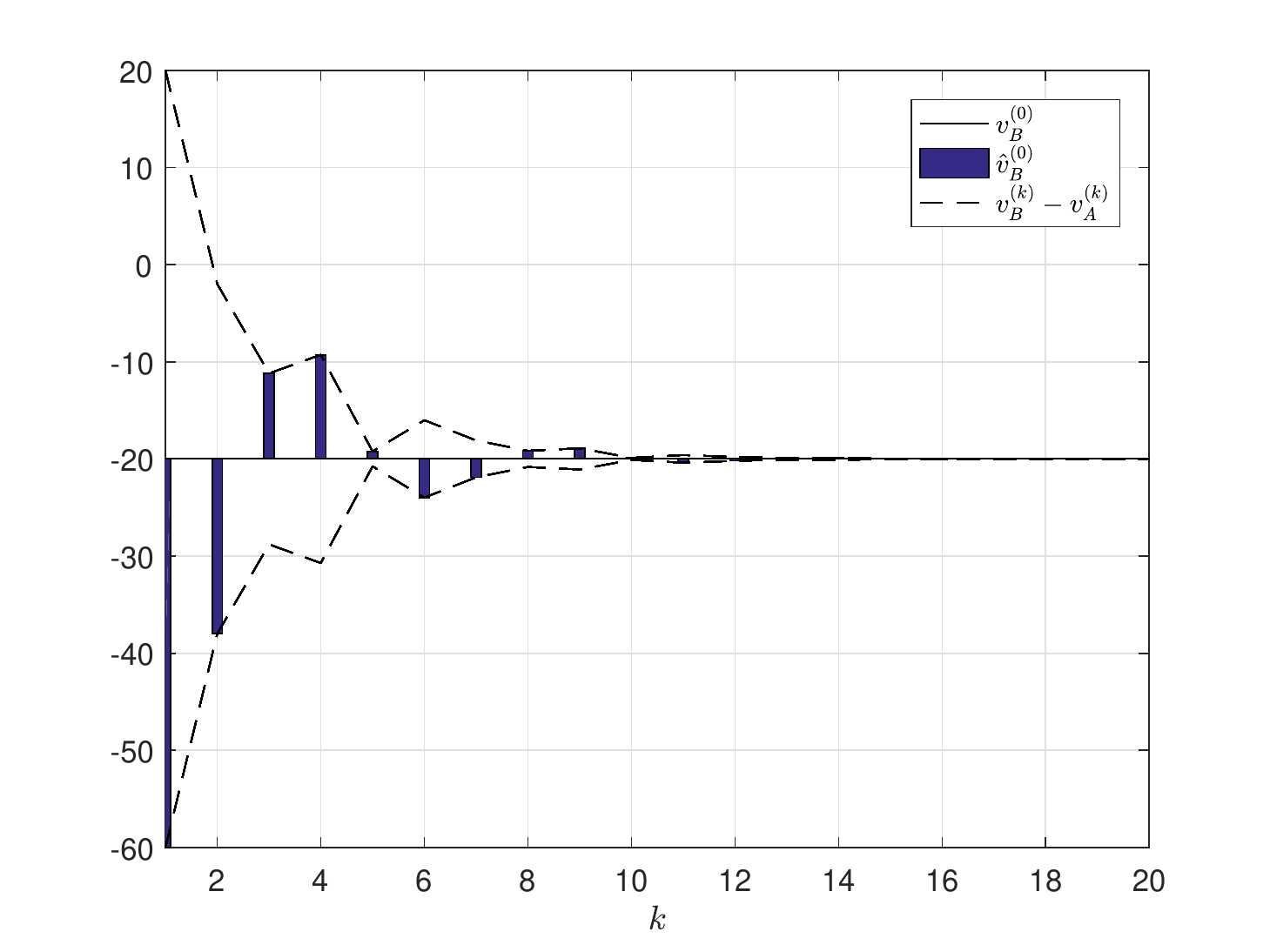}
			\subcaption{The initial velocity of agent B and  its associated estimate computed by agent A.}
		\end{minipage}
		\caption{All four agents' position and velocity trajectories and estimation of agent B's initial states calculated by agent A in a fast consensus network.}\label{fig-EstimationError}
	\end{figure}

Fig.~\ref{fig-EstimationError}a and Fig. 3c exhibit the position and velocity trajectories of all four agents, accordingly. For the case in which agents A and B have complete knowledge about $a^{(k)}_{AB}$, i.e. no decoupling is applied, the initial position and velocity components associated with agent B can be estimated by agent A. This is  illustrated in Fig. 3b and Fig 3d. One can observe that   $\hat{p}^{(0)}_B\to p^{(0)}_B$ and $\hat{v}^{(0)}_B\to v^{(0)}_B$ as $k\to\infty$.  One should note that at even though at $k=5$  the consensus is not reached but since local $\delta$-agreement is achieved between agents A and B for a small $\delta$, the estimated values of agent B's initial states are accurate.
	
\begin{figure}[htb]
		\begin{minipage}{0.45\textwidth}
			\centering
			\includegraphics[scale=.45]{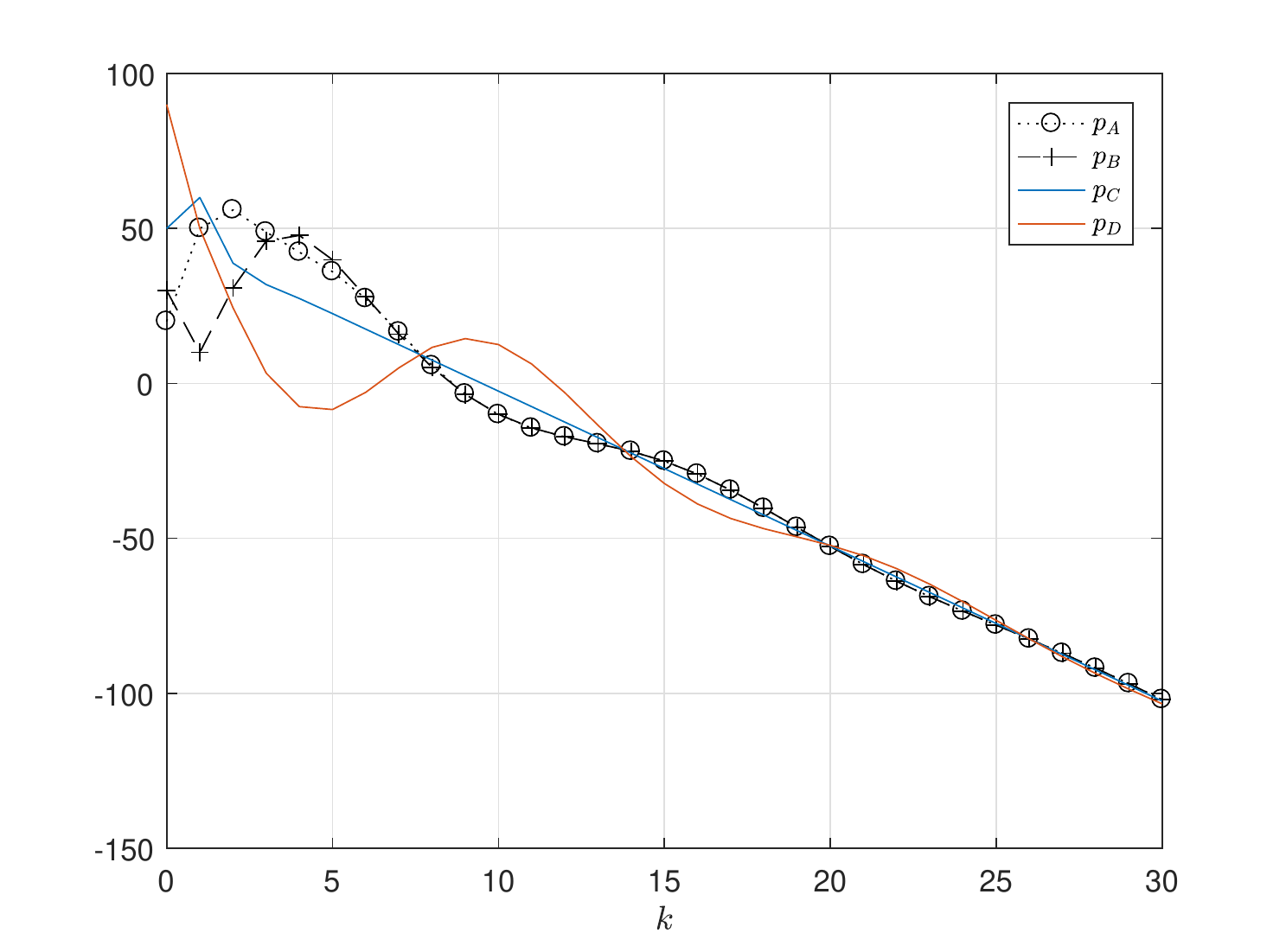}
			\subcaption{Position trajectories of all four agents.}
		\end{minipage}
		\begin{minipage}{0.45\textwidth}
			\centering
			\includegraphics[scale=.45]{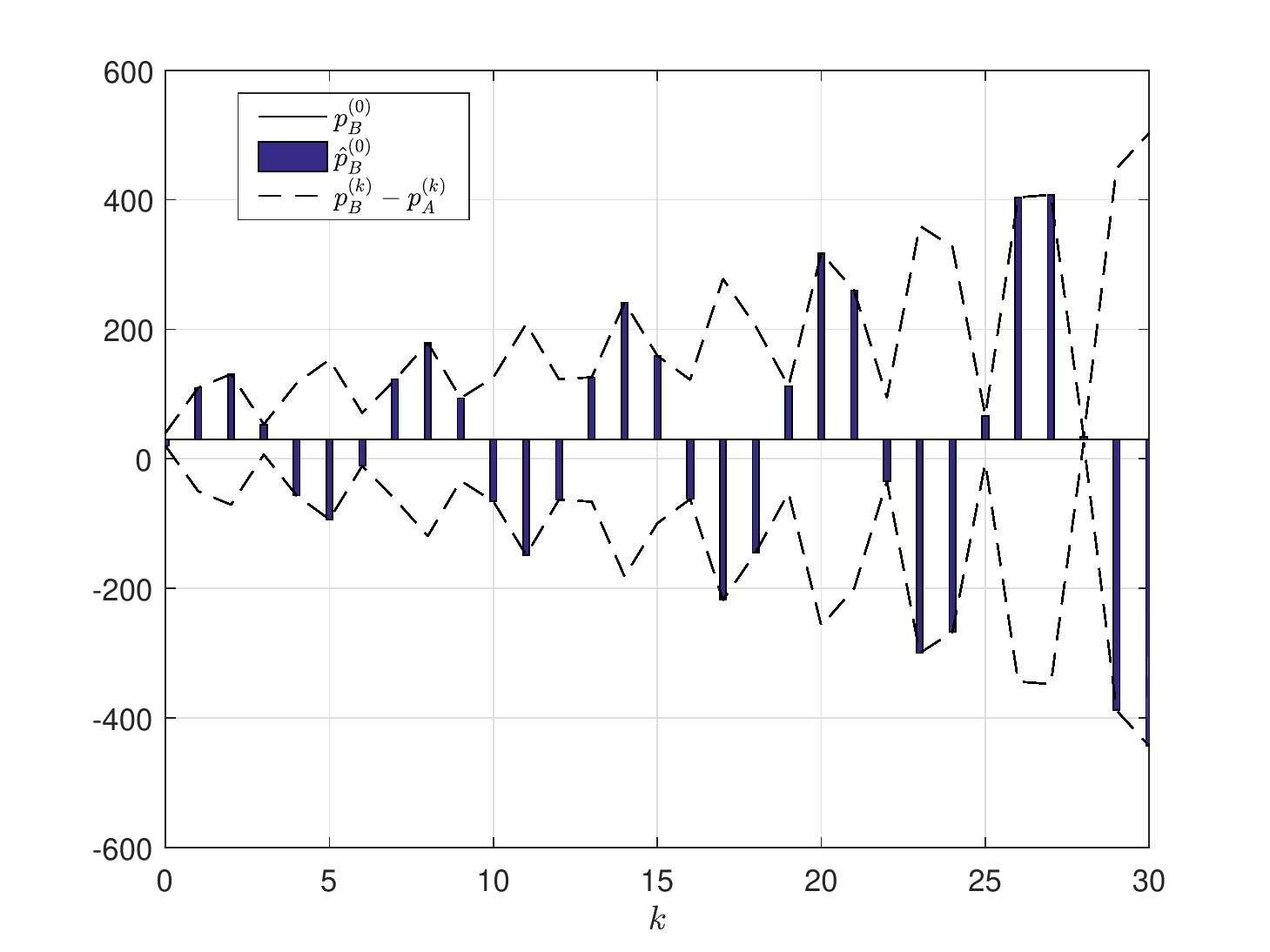}
			\subcaption{The initial position of agent B and its associated estimates computed by agent A.}
		\end{minipage}
		\begin{minipage}{0.45\textwidth}
			\centering
			\includegraphics[scale=.45]{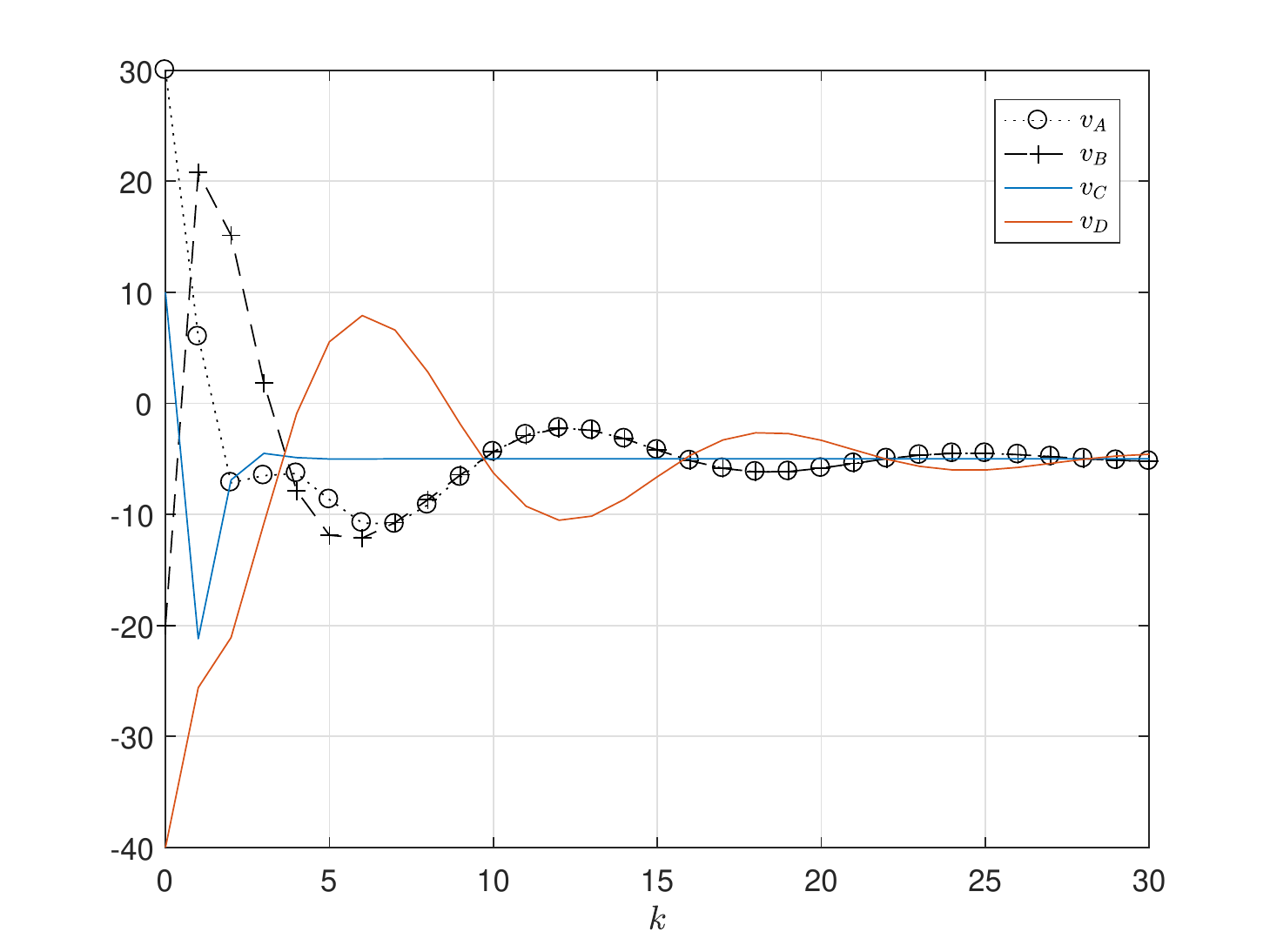}
			\subcaption{Velocity trajectories of all four agents.}
		\end{minipage}
		\begin{minipage}{0.45\textwidth}
			\centering
			\includegraphics[scale=.45]{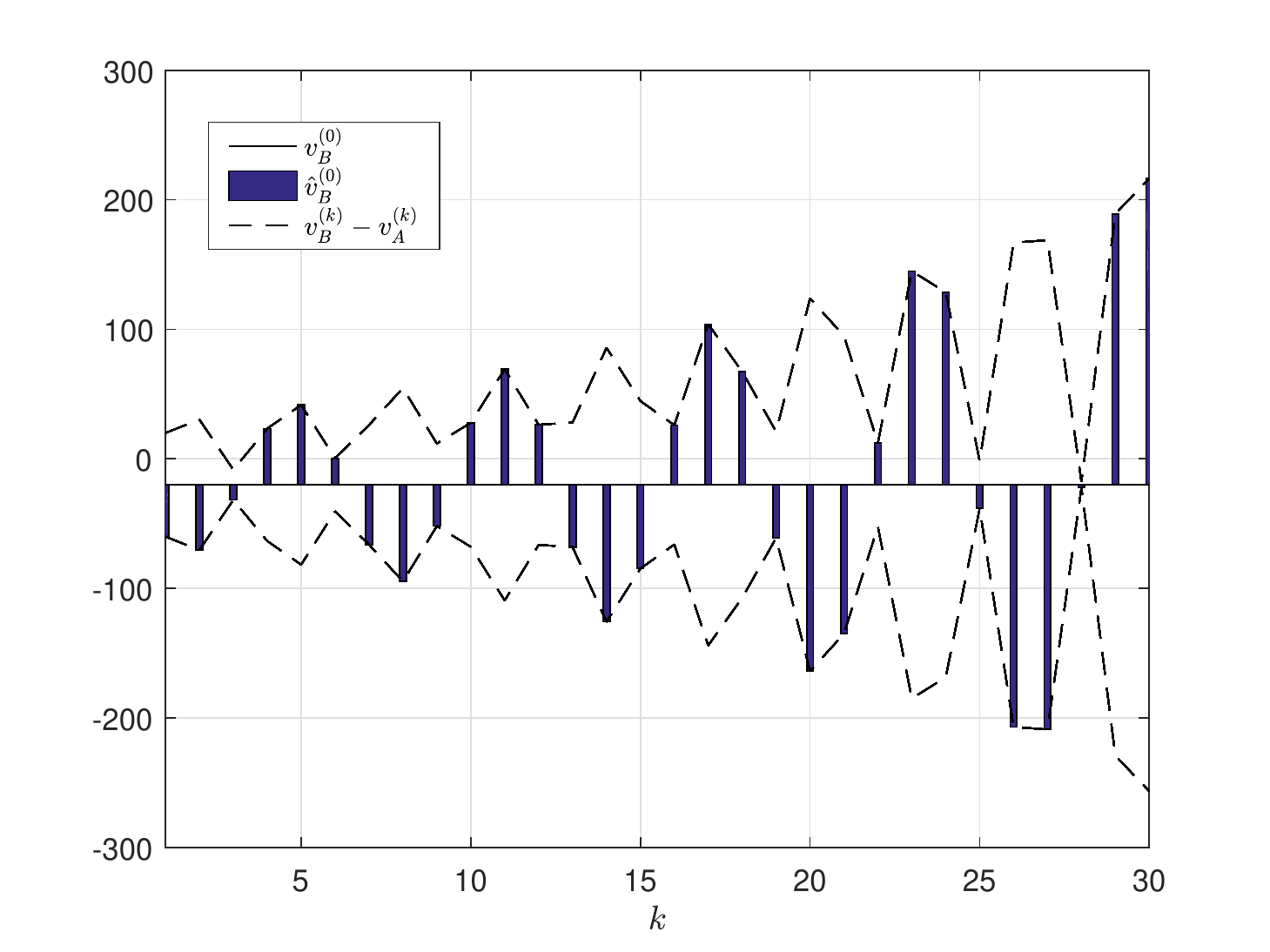}
			\subcaption{The initial velocity of agent B and  its associated estimate computed by agent A.}
		\end{minipage}
		\caption{All four agents' position and velocity trajectories and estimation of agent B's initial states calculated by agent A in a slow consensus network.}\label{fig-EstimationError-Divergent}
	\end{figure}

	As expected by Theorem~\ref{thm-DisclosureWithoutDecouple}, in Fig. \ref{fig-EstimationError}, 
	agent A estimates B's initial states with  error values which converge to $0$ as time $k\to\infty$.
	It occurs in a consensus process with a sufficiently fast convergence rate.
	Next, with all the coupling weights $a_{ij}$s multiplied by a factor $0.8$, the consensus
	convergence rate reduces accordingly. In this case, estimation of B's initial states is inaccurate
	as illustrated in Fig.~\ref{fig-EstimationError-Divergent}. That is, B's privacy still remains intact.
	
		  Finally,  as far as the computation complexity of the proposed scheme is concerned, it was recorded  that 
		the average computation time for each step of simulation is $174.9$~ms on a laptop with 2.9 GHz Intel Core i7 Dell laptop using Matlab R2017a, which runs  Linux on  a Virtual Machine. Each step contains $12$ encryption processes and $8$ decryption processes   associated with  all $4$ agents.

	\section{Conclusion}\label{sec-Conclusion}

	In this paper,  we have provided a cryptography-based secure   consensus protocol
		for networks of second-order agents.  The proposed protocol achieved consensus asymptotically among all agents while keeping the privacy of individuals intact through    decoupling and varying the communication weights within some admissible range. In addition, for the case which the communication weights information is available to all agents, we have performed a full privacy analysis and shown that violation of individual's privacy depends on the network convergence rate.

%

	\bibliographystyle{IEEEtran}
	\bibliography{PrivateConsensus}%

\end{document}